\documentclass[
  twocolumn,aps,prx,showpacs,amsmath,longbibliography,
  amssymb,superscriptaddress,bibnotes]{revtex4-1} 

\usepackage{hyperref}
\usepackage{graphicx}
\usepackage{mathtools}
\usepackage{hyperref}
\usepackage{color}

\usepackage[export]{adjustbox}

\newcommand{\tmax}{t_{\textrm{max}}}
\newcommand{\mbf}[1]{\mathbf{#1}}

\newcommand{\Ra}{i}
\newcommand{\Rb}{j}

\newcommand{\na}{\gamma}
\newcommand{\nb}{\nu}

\newcommand{\ba}{b}
\newcommand{\bc}{b^\dagger}
\newcommand{\Ba}{\mbf{b}}
\newcommand{\Bc}{\mbf{b}^\dagger}

\newcommand{\dBa}{\delta\mbf{b}}
\newcommand{\dBc}{\delta\mbf{b}^\dagger}

\newcommand{\Pa}{\mbf{\Phi}}
\newcommand{\Pc}{\mbf{\Phi}^\dagger}

\newcommand{\Pceff}{\mbf{\Phi}^\dagger_{\textrm{eff}}}

\newcommand{\n}{\hat{n}}
\newcommand{\C}{\mathcal{C}}

\newcommand{\Simp}{\mathcal{S}_{\textrm{imp}}}
\newcommand{\D}{\mbf{\Delta}}
\newcommand{\Teff}{T_{\textrm{eff}}}
\newcommand{\Nmax}{N_{\textrm{max}}}
\newcommand{\Tr}{\textrm{Tr}}

\newcommand{\pt}{\partial_t}

\newcommand{\Sp}{\hat{\Sigma}}
\newcommand{\Gp}{\hat{G}}
\newcommand{\Hloc}{\hat{H}}
\newcommand{\astcycl}{\mathrlap{\kern0.085em{\circlearrowright}}\ast}

\newcommand{\Gk}{\mbf{G}_{\mbf{k}}}
\newcommand{\Gl}{\mbf{G}_{L}}

\newcommand{\Gc}{\mbf{G}}

\newcommand{\Ucdyn}{U_{c}^{\textrm{dyn}}}
\newcommand{\CC}{{|\phi|_c}}
\newcommand{\AM}{{|\phi|_{AM}}}

\newcommand{\Z}{\mathcal{Z}}

\newcommand{\taucycl}{\mathrlap{\kern0.42em{\bullet}}\circlearrowright}

\newcommand{\ttau}{\urcorner}

\newcommand{\bba}{\mbf{b}^{\phantom{\dagger}}}
\newcommand{\bbc}{\mbf{b}^{\dagger}}

\newcommand{\ec}{\eta^{\dagger}}

\newcommand{\bea}{\boldsymbol{\eta}^{\phantom{\dagger}}}
\newcommand{\bec}{\boldsymbol{\eta}^{\dagger}}

\newcommand{\bG}{\mbf{G}}
\newcommand{\bPhi}{\boldsymbol\Phi}
\newcommand{\bDelta}{\boldsymbol{\Delta}}

\newcommand{\la}{\langle}
\newcommand{\ra}{\rangle}

\newcommand{\TC}{\mathcal{T}_\C}

\newcommand{\ct}{t}

\begin{document}

\title{Nonequilibrium dynamical mean-field theory 
for bosonic lattice models}

\author{Hugo U.~R.~Strand}
\email{hugo.strand@unifr.ch}
\affiliation{Department of Physics, University of Fribourg, 1700 Fribourg, Switzerland} 

\author{Martin Eckstein}
\affiliation{Max Planck Research Department for Structural Dynamics, University of Hamburg-CFEL, Hamburg, Germany} 

\author{Philipp Werner}
\email{philipp.werner@unifr.ch}
\affiliation{Department of Physics, University of Fribourg, 1700 Fribourg, Switzerland} 

\date{\today} 


\begin{abstract}
We develop the nonequilibrium extension of bosonic dynamical mean field theory (BDMFT) and a Nambu real-time strong-coupling perturbative impurity solver.
In contrast to Gutzwiller mean-field theory and strong coupling perturbative approaches, nonequilibrium BDMFT captures not only dynamical transitions, but also damping and thermalization effects at finite temperature.
We apply the formalism to quenches in the Bose-Hubbard model, starting both from the normal and Bose-condensed phases.
Depending on the parameter regime, one observes qualitatively different dynamical properties, such as rapid thermalization, trapping in metastable superfluid or normal states, as well as long-lived or strongly damped amplitude oscillations. 
We summarize our results in non-equilibrium ``phase diagrams'' which map out the different dynamical regimes.
\end{abstract}

\maketitle
\makeatletter
\let\toc@pre\relax
\let\toc@post\relax
\makeatother

\section{Introduction}


Cold atomic gases trapped in an optical lattice provide a unique play-ground to explore equilibrium and nonequilibrium properties of interacting many-particle systems \cite{Morsch:2006vn, Bloch:2008uq}. They enable an almost ideal realization of the low-energy effective Hamiltonians (the fermionic and bosonic Hubbard models \cite{Hubbard:1963aa, Fisher:1989kl}) which have been studied in the condensed matter context for a long time, and whose properties are still not yet fully understood. A big advantage of cold atoms, as compared to condensed matter systems, is that interaction parameters can be tuned almost arbitrarily, and that the lattice spacings and characteristic time-scales are much larger \cite{Jaksch:1998vn}. For bosonic atoms, the Mott insulating and superfluid regime can easily be accessed \cite{Morsch:2006vn} and the experimental control is so precise that the use of cold atoms as ``quantum simulators" becomes a realistic option \cite{Trotzky:2010fk} (for a recent review see Ref.\ \onlinecite{Kennett:2013fk}).


A particularly interesting aspect of cold atom experi\-ments is the possibility to study the time-evolution of interacting many-body systems \cite{Greiner:2002fk,Sebby-Strabley:2007vn, Will:2010uq, Bakr:2010dq, Bissbort:2011bs, Endres:2012bs, Cheneau:2012ys, Trotzky:2012kx, Braun:2014kx}. This was beautifully demonstrated in the seminal work by Greiner {\it et al.~}\cite{Greiner:2002fk}, who measured the condensate collapse-and-revival oscillations after a quench in a Bose-Hubbard system from the superfluid to the Mott regime. In contrast to equilibrium, where the phase diagram and correlation functions of the Bose-Hubbard model \cite{Capogrosso-Sansone:2007lh} can be computed accurately using Monte Carlo simulations \cite{Pollet:2012ly}, the real-time evolution of interacting bosonic lattice systems is a big computational challenge. 




In one dimension, density matrix renormalization group (DMRG) methods \cite{Schollwock:2005ly} can be used to simulate the time-evolution after a quench on relatively large lattices, but a rapid entanglement growth limits the accessible time-scale \cite{Trotzky:2012kx}. Still, DMRG calculations have provided important insights into the short time dynamics, as measured in 1D optical lattices \cite{Cheneau:2012ys, Trotzky:2012kx, Braun:2014kx}. 
Kollath {\it et al.~}\cite{Kollath:2007ys} used non-local correlators to study relaxation and thermalization. They showed that an initially superfluid system is trapped in a nonthermal steady state after quenching the interaction deep into the Mott regime, while thermalization occurs after quenches to intermediate interactions.
%
Also the eigenstate thermalization hypothesis has been explored \cite{Roux:2009ys, Sorg:2014vn} and debated \cite{Rigol:2010zr, Roux:2010ly} in this context.
%
%
%
A more recent development is the time-dependent variational Monte Carlo (tVMC) approach that shows good agreement with DMRG in 1D without being limited in time \cite{Carleo:2012zr}. It has also been applied to 2D systems and is not inherently limited to any dimensionality \cite{Carleo:2014ly}. While tVMC is well suited for studying the spread of correlations, it is a method that treats finite systems, 
which complicates the study of thermalization \cite{Biroli:2010qf}.


In three dimensions, perturbation theory \cite{Trefzger:2011uq, Kennett:2011fk, Dutta:2012kx, Dutta:2014uq}, and Gutzwiller mean-field (MF)  \cite{Huber:2007ys, Wolf:2010bh, Sciolla:2010uq, Sciolla:2011kx, Snoek:2011hc, Krutitsky:2011qf} calculations have been performed. Both work in specific regions of the phase diagram, but generally fail to describe finite temperature relaxation and thermalization phenomena.
Hence, while being accessible experimentally \cite{Braun:2014kx}, out-of-equilibrium phenomena in the three dimensional Bose-Hubbard model remain largely unexplored \cite{Trotzky:2012kx, Cheneau:2012ys, Braun:2014kx} from the theoretical point of view. 
Describing the generic relaxation phenomena and nonthermal transient states, as well as mapping out the different dynamical regimes of this model is fundamental to our understanding of nonequilibrium lattice bosons.
A clear picture of the nonequilibrium properties of the homogeneous bulk-system is also important for the interpretation of more complicated experimental set-ups.  
%
For example, one open question is whether damped superfluid collapse-and-revival oscillations are a dynamical feature of the homogeneous system, or an effect of the trapping potential or other processes not considered in the Bose-Hubbard description \cite{Greiner:2002fk, Will:2010uq}.

%


A computationally tractable and promising scheme, which allows to address such issues, is the nonequilibrium generalization of bosonic dynamical mean field theory (BDMFT).
This method is formulated in the thermodynamical limit, and thus enables the study of relaxation and thermalization phenomena in infinite systems \cite{Aoki:2014kx}.
The equilibrium version of BDMFT \cite{Byczuk:2008nx, Hubener:2009cr, Anders:2010uq, Anders:2011uq} produces phase diagrams, condensate fractions, and correlation functions with remarkable accuracy \cite{Anders:2011uq}.
While the extension of this formalism to nonequilibrium situations is analogous to the fermionic case \cite{Aoki:2014kx}, and essentially involves the replacement of the imaginary-time interval by a Kadanoff-Baym contour, there are a number of practical challenges.
The most important one is the development of a suitable bosonic impurity solver. The exact continuous-time quantum Monte Carlo (CT-QMC) impurity solver of Ref.~\onlinecite{Anders:2010uq} cannot easily be applied to nonequilibrium problems, because of a dynamical sign problem \cite{Werner:2009tg}, while exact diagonalization based solvers are even more limited than in the fermionic case \cite{Gramsch:2013fk}, due to the larger local Hilbert space. Weak-coupling perturbation theory is not an option if one is interested in Mott physics. Instead, we will develop and benchmark an impurity solver based on the lowest order strong-coupling perturbation theory, i.e.\ the non-crossing approximation (NCA) \cite{Keiter:1971hc}.
As a first application of this new scheme, we will map out the different dynamical regimes of both the symmetric and symmetry broken states, searching for thermalization and trapping phenomena after a quench of the interaction parameter.


This paper is organized as follows: 
In section \ref{sec:Theory} we give an overview of the Bose-Hubbard model, the nonequilibrium generalization of BDMFT [Sec.\ \ref{sec:BDMFT}], the NCA impurity solver [Sec.\ \ref{sec:NCA}], the energy calculations [Sec.\ \ref{sec:Energy}], and our numerical implementation [Sec.\ \ref{sec:Numerics}]. In Sec.\ \ref{sec:Results} we first present benchmark calculations showing density and energy conservation and discuss the lowest order spectral moments [Sec.\ \ref{sec:Benchmark}]. The dynamical regimes in the normal phase are mapped out in Sec.\ \ref{sec:Mott}. In Sec.\ \ref{sec:SuperFluid} we consider superfluid initial states, and after an overview of the relaxation regimes in Sec.\ \ref{sec:SuperFluidRegimes}, we study the dynamics for short times in Sec.\ \ref{sec:ShortTimeDynamics}, and long times in Sec.\ \ref{sec:LongTimeDynamics}. The findings are summarized in Sec.\ \ref{sec:SuperFluidPhaseDiagram} in the form of a nonequilibrium ``phase diagram''. Sec.\ \ref{sec:Conclusions} is devoted to conclusions. We also provide a derivation of nonequilibrium BDMFT in Appendix \ref{app:BDMFT}, and discuss the details of the Nambu generalization of NCA in Appendix \ref{app:NCA}.

\section{Theory} \label{sec:Theory}


We consider the simplest model for bosonic atoms in an optical lattice, namely the Bose-Hubbard model \cite{Fisher:1989kl, Jaksch:1998vn}
\begin{equation}
  H = -J \sum_{\langle \Ra,\Rb \rangle} (\bc_\Ra \ba_\Rb + \bc_\Rb \ba_\Ra)
  +\frac{U}{2} \sum_\Ra \n_\Ra(\n_\Ra-1) - \mu \sum_\Ra \n_\Ra
  \, , \label{eq:Model}
\end{equation}
 where $\bc_\Ra$ ($\ba_\Ra$) and $\n_\Ra$ are the bosonic creation (annihilation) and number operators acting on site $\Ra$, $\mu$ is the chemical potential, and $U$ the local pair interaction which competes with the nearest neighbor hopping $J$ that we take as our unit of energy.

\subsection{Nonequilibrium bosonic dynamical mean-field theory} \label{sec:BDMFT}


By extending the equilibrium bosonic dynamical mean field theory (BDMFT) \cite{Anders:2010uq, Anders:2011uq} to the three-branch Kadanoff-Baym contour $\mathcal{C}$ ($0 \! \rightarrow \! t_\text{max} \! \rightarrow \! 0 \! \rightarrow \! -i\beta$) \cite{Aoki:2014kx, Stefanucci:2013oq}, we obtain the bosonic impurity action
\begin{align}
  \Simp & =
  \int_\C dt\,\Big( -\mu(t)\n(t) + \frac{U}{2}\n(t)(\n(t)-1) \Big) 
  \label{eq:Seff}\\
  &-\int_\C dt \, \Pceff(t)\Ba(t) 
  + \frac{1}{2} \iint_\C dt\,dt'\, \Bc(t)\D(t, t')\Ba(t') 
  \, , \nonumber
\end{align}
 where $\Bc$ is the Nambu spinor $\Bc = (\bc, \ba)$, $\D(t, t')$ the hybridization function, and $\Pceff$ the effective symmetry breaking field, 
which is defined in terms of $\D$, the local condensate fraction $\Pc = (\phi^*, \phi)$, and the lattice coordination number $z$ as 
\begin{equation}
  \Pceff(t) =  zJ\Pc(t) + \int_\C dt'\, \Pc(t')\D(t', t) \, . \label{phieff}
\end{equation}
For a detailed derivation of this action, see App.\ \ref{app:BDMFT}. Note that the single-particle fluctuations (the $\D$ term in Eq.\ (\ref{eq:Seff}) and Eq.\ (\ref{phieff})) enter as a correction to the mean-field action \cite{Sachdev:1999fk}, which would be obtained by taking the infinite dimensional limit $z \rightarrow \infty$ at fixed $zJ$ (or analogously $\D \rightarrow 0$) \cite{Anders:2011uq}. 


The solution of the impurity model yields the connected impurity Green's function
\begin{equation*}
  \Gc(t,t') = 
  -i\langle \mathcal{T}_\C \Ba(t) \Bc(t')\rangle + i\Pa(t)\Pc(t') \, ,
\end{equation*}
where $\mathcal{T}_\mathcal{C}$ is the time-ordering operator on the contour $\mathcal{C}$, and the local condensate fraction is
\begin{equation*}
  \Pa(t)=\langle \Ba(t) \rangle \, .
\end{equation*}
The BDMFT self-consistency loop is closed by computing the lattice Green's function $\Gk$ from $\Gc$, and then expressing the hybridization function $\D$ in terms of the local lattice Green's function $\Gl = \frac{1}{N_{\mbf{k}}}\sum_{\mbf{k}} \Gk$ (at self consistency $\Gl = \Gc$) \cite{Aoki:2014kx}.
In the present study we employ the simplified self-consistency relation 
\begin{equation*}
  \D(t, t') = (3J)^2 \Gc(t, t') \, ,
\end{equation*}
and set $z=6$, which corresponds to a non-interacting semi-circular density of states (DOS) with the same bandwidth $W = 12J$ and lattice coordination number $z$ as the 3D cubic lattice with nearest neighbor hopping $J$.


\subsection{Non-crossing approximation impurity solver}
\label{sec:NCA}

The previous BDMFT equilibrium studies employed a hybridization expansion CT-QMC impurity solver \cite{Anders:2010uq, Anders:2011uq}. However, the extension of this technique to the contour-action in Eq.\ (\ref{eq:Seff}) does not look promising, because the dynamical sign problem from the expansion along the real-time branches \cite{Werner:2010pb} will add to the inherent sign problem of the hybridization expansion (in the superfluid regime).
We therefore solve the BDMFT effective impurity action using the first order self-consistent strong coupling expansion.
The generalization of strong coupling expansions to real-time impurity problems has been presented in Ref.~\cite{Eckstein:2010fk}. To treat the BDMFT effective action in Eq.\ (\ref{eq:Seff}) we have generalized the formalism to systems with symmetry breaking, as discussed in App.~\ref{app:NCA}.

In short we follow the standard procedure and introduce pseudo-particle second quantization operators $p^{\phantom{\dagger}}_\Gamma$ and $p^\dagger_\Gamma$ for each local occupation number many-body state $| \Gamma \rangle$.
This maps the local Hamiltonian to a quadratic term $\sum_{\Gamma\Gamma'} \Hloc(t)_{\Gamma\Gamma'} p^\dagger_\Gamma p^{\phantom{\dagger}}_{\Gamma'}$, while the hybridization $\bDelta$ turns into a pseudo-particle interaction.
Expanding to first order in $\bDelta$ gives the NCA of Ref.\ \onlinecite{Eckstein:2010fk} generalized to Nambu formalism.


\begin{figure}
  \begin{flushleft}
    \includegraphics[scale=1] {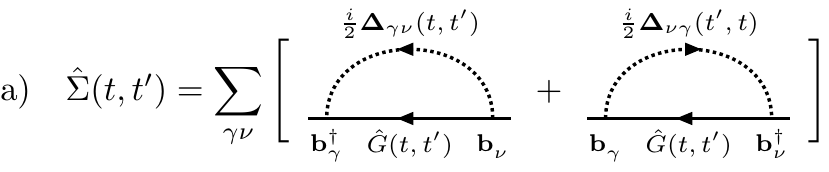} \\
    \includegraphics[scale=1] {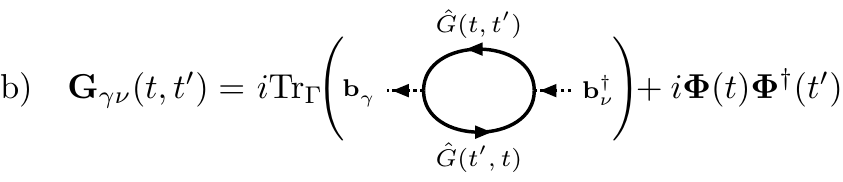}
  \end{flushleft}
\caption{\label{fig:NCA} 
 NCA diagram representations of a) the pseudo-particle self-energy $\Sp$, and b) the single-particle Green's function $\Gc_{\na \nb}$.
} \end{figure}

The corresponding NCA pseudo-particle self-energy $\Sp = \Sp_{\Gamma\Gamma'}$ consists of the two shell diagrams with a directed hybridization line (see Fig.~\ref{fig:NCA} and App.\ \ref{app:NCASigma})
\begin{align}
  \Sp(t, t') = 
  \frac{i}{2} \sum_{\na \nb} \Big( & 
  \D_{\na \nb}(t, t') \left[ \Bc_\na \Gp(t, t') \Ba_\nb \right] + \nonumber \\ & 
  \D_{\nb \na}(t', t) \left[ \Ba_\na \Gp(t, t') \Bc_\nb \right] \Big) \, ,
  \label{eq:NCASigma}
\end{align}
where $\Gp = \Gp_{\Gamma\Gamma'}$ is the pseudo-particle Green's function, $\gamma$ and $\nu$ are Nambu indices, and $\Ba_\na$ is the tensor $(\Ba_\na)_{\Gamma\Gamma'} = \langle \Gamma | \Ba_\na | \Gamma' \rangle$ (operator products are implicit matrix products).
The pseudo-particle Dyson equation takes the form 
\begin{equation*}
( i \pt + \Hloc(t) )\Gp  - \mbox{$\Sp \astcycl \Gp$} = 0,
\end{equation*}
where $\Hloc(t)$ is the static part in Eq.\ (\ref{eq:Seff}), $\Hloc(t) = U(t)(\n^2 - \n)/2 - \mu(t)\n - \Pceff(t) \Ba$, and  $\mbox{$\Sp \astcycl \Gp$}$ denotes cyclic convolution on $\C$, $(\mbox{$\Sp \astcycl \Gp$})(t, t') = \int_{t' \prec \bar{t} \prec t} d\bar{t} \, \Sp(t, \bar{t}) \, \Gp(\bar{t}, t')$ \cite{Eckstein:2010fk}.

Within NCA, $\Gp$ and $\Sp$ are calculated self-consistently, and local observables are determined from the reduced local density matrix $\hat{\rho}(t) = i\Gp^<(t,t)$, yielding the local condensate as 
\begin{equation*}
 \Pa_\na(t) = 
 \langle \Ba_\na(t) \rangle = \Tr_\Gamma [ \Ba_\na \hat{\rho}(t) ],
\end{equation*}
while response functions must be determined diagrammatically. In particular, the connected single-particle impurity Green's function $\Gc$ is obtained from the bubble diagram without hybridization insertions
(see Fig.\ \ref{fig:NCA} and App.\ \ref{app:NCAGsp})
\begin{multline}
  \Gc_{\na \nb}(t, t') = \\
  i \Tr_\Gamma \left[ \Gp(t', t) \Ba_\na \, \Gp(t, t') \Bc_\nb \right]
  + i \Pa_\na(t) \Pc_\nb(t') .
  \label{eq:NCAGsp}
\end{multline}

\subsection{Total energy components} \label{sec:Energy}


The total energy $E_t$ of the system, is the sum of the (connected) kinetic energy $E_k$, the condensate energy (or disconnected kinetic energy) $E_c$, and the local interaction energy $E_i$, $E_t = E_k + E_c + E_i$. 
Using $\Gc$ and $\D$, $E_k$ is given by \cite{Aoki:2014kx}
\begin{equation*}
  E_k(t) = \frac{i}{2} \Tr \left[ (\D * \Gc)^<(t, t) \right],
\end{equation*}
 $E_c$ depends on $\phi(t) = \langle \ba(t) \rangle = \Tr_\Gamma[ \ba \, \hat{\rho}(t) ]$ as 
\begin{equation*}
  E_c(t) = -zJ(t) | \phi(t) |^2,
\end{equation*}
and $E_i$ can be written in terms of $\langle \n^2 \rangle(t) = \Tr_\Gamma[ \n^2 \, \hat{\rho}(t) ]$ and $\langle \n \rangle(t) = \Tr_\Gamma[ \n \, \hat{\rho}(t) ]$ as
\begin{equation*}
  E_i(t) = U(t) ( \langle \n^2 \rangle(t) - \langle \n \rangle(t) )/2.
\end{equation*}

\subsection{Numerical implementation} \label{sec:Numerics}

We solve the pseudo-particle Dyson equation using a fifth order multi-step method \cite{Brunner:1986ff, Eckstein:2010fk} on an uniformly discretized time grid.
To ensure negligible real-time discretization errors we monitor the total energy and density, which both are constants of motion of the conserving NCA \cite{Eckstein:2010fk} (the gauge property $\mu \rightarrow \mu + \delta\mu(t) \Rightarrow \ba \rightarrow \ba e^{-i \int_0^t d\bar{t} \, \delta\mu(\bar{t})}$ ensures $\partial_t \langle \n \rangle = 0$).
In principle the local Fock space is unbounded, but for $U>0$ it can safely be truncated, keeping only $\Nmax$ states. The cut-off error is controlled by monitoring the drift in $\Tr_\Gamma[ \hat{\rho} ]$ away from unity. Close to the $\langle \n \rangle = 1$ superfluid transition at $U \gtrsim J$, the results are converged for $\Nmax = 5$ to $11$.

The computational limitations of our real-time BDMFT+NCA implementation are very similar to the real-time fermionic DMFT+NCA case \cite{Eckstein:2010fk}. Memory is \emph{the} limiting factor when working with two time response functions, whose storage size scales quadratically with the number of time steps. The local Fock space in the bosonic case adds one or two orders of magnitude in memory usage, compared to the single-band fermionic case. A further limitation is the quadratic energy dependence of the local occupation number states $| \Gamma \rangle$, scaling with $E_\Gamma \sim \la \Gamma | U\n^2 | \Gamma \ra = Un_\Gamma^2$. This induces a pseudo-particle time dependence $\Gp_{\Gamma \Gamma}(t, t') \sim e^{-iUn_\Gamma^2(t-t')}$, which means that including higher occupation number states, by increasing $\Nmax$, also requires a finer time discretization.

\section{Results} \label{sec:Results}


\begin{figure}
\includegraphics[scale=1.0]{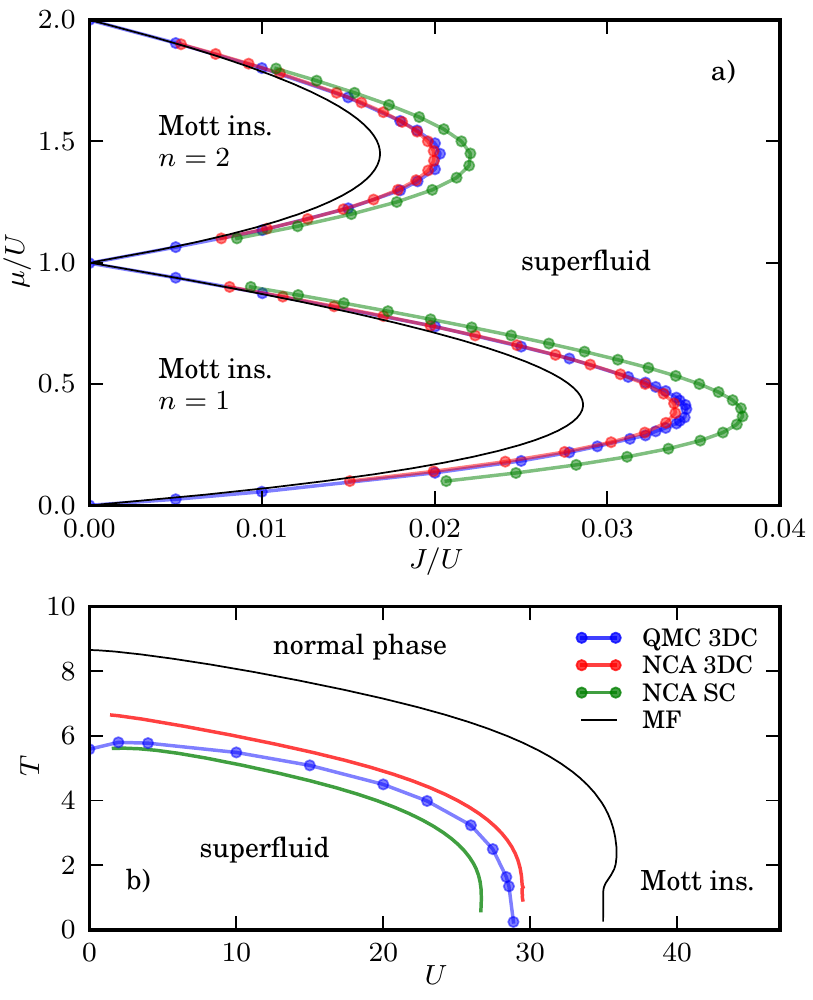}
\caption{\label{fig:PhaseDiag} (color online) BDMFT superfluid phase boundary for CT-QMC \cite{Anders:2011uq} (blue) and NCA (red) on the 3D cubic lattice, NCA with a semi circular DOS (green), and mean-field theory (black). Panel a) shows the ($J/U$, $\mu/U$) plane at $T=1.5$, and panel b) the ($U$, $T$) plane for $\langle \n \rangle = 1$.}
\end{figure}

\subsection{Benchmark calculations} \label{sec:Benchmark}

Even though BDMFT neglects spatial fluctuations, the equilibrium results for the 3D Bose-Hubbard model are in good quantitative agreement \cite{Anders:2010uq, Anders:2011uq} with high precision lattice QMC calculations \cite{Capogrosso-Sansone:2007lh} and high-order perturbation theory \cite{Teichmann:2009bh}, for both the phase diagram and local correlation functions.
For example, the critical couplings at the $\langle \n \rangle = 1$ superfluid-Mott transition are $(J/U)_c = 0.0345 \pm 0.0004$ (BDMFT at $\beta J = 2$), and $(J/U)_c = 0.03408(2)$ (lattice QMC), see Ref.\ \onlinecite{Anders:2011uq} for an explicit comparison of phase diagrams.

To assess the validity of the NCA approximation we compare its superfluid phase boundary for the 3D cubic lattice with the (within BDMFT) exact CT-QMC result, see Fig.\ \ref{fig:PhaseDiag}. It is evident that already this lowest order strong coupling expansion provides a very good approximation with $(J/U)_c \approx 0.0340$ (at $T=1.5$), as expected, considering the success of the linked cluster expansion \cite{Kauch:2012ve}. The simplified self-consistency based on the semi-circular DOS leads to a shift in the phase boundaries (Fig.\ \ref{fig:PhaseDiag}) with $(J/U)_c \approx 0.0378$, but we expect that the qualitative features of the solution, both in and out of equilibrium, remain unchanged.

Note that the Mott phase is only present at integer fillings. Hence, in order to study quenches between the superfluid and Mott insulator, we limit our calculations to $\langle \n \rangle = 1$. Strictly speaking the Mott insulator exist only at zero temperature, with a smooth crossover to the normal phase, see Fig.\ \ref{fig:PhaseDiag}. However we follow Ref.\ \onlinecite{Capogrosso-Sansone:2007lh} and define the \emph{Mott regime} as the whole region $U > U_c(T=0)$, where the low temperature superfluid phase is absent.


For instantaneous interaction quenches the final total energy $E^{(f)}_{t}$ is given by the initial equilibrium total energy $E^{(i)}_{t}$ and an additional interaction energy contribution $E^{(f)}_{t} = E^{(i)}_{t} + (U_f/U_i - 1) E^{(i)}_{i}$ (due to the sudden change of $U$ from $U_i$ to $U_f$ at $t=0$). Given $E^{(f)}_{t}$ and $U_f$ the effective temperature $\Teff$ of the system after thermalization can be determined using separate equilibrium calculations. 
The resulting non-equilibrium $(U_{f}, \Teff)$ pair of a quench can be used to determine the final state after eventual thermalization by direct comparison with the equilibrium $(U, T)$ phase boundaries. This will be used throughout this study in order to produce combined equilibrium $(U, T)$ and non-equilibrium $(U_f, \Teff)$  ``phase diagrams''.


\begin{figure}
\includegraphics[scale=1]{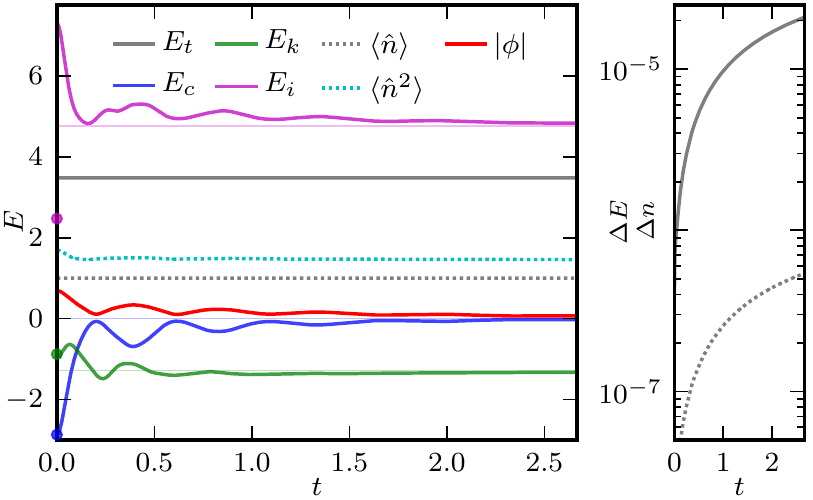}
\caption{\label{fig:SFquenchEnergies} (color online) Time-evolution of energies and observables (lines), and thermal values (thin lines) for the 
superfluid to normal phase quench from $U_i=6$ ($T_i=4.5$) to $U_f=21$ ($\Teff \approx 9.81$) (left panel), 
and time discretization induced drifts $\Delta n = \langle \n(t) \rangle - \langle \n(0) \rangle$ and $\Delta E_{t}(t) = E_{t}(t) - E_{t}(0)$ (right panel).} \label{fig_energies}
\end{figure}


\begin{figure*}
\includegraphics[scale=1]
{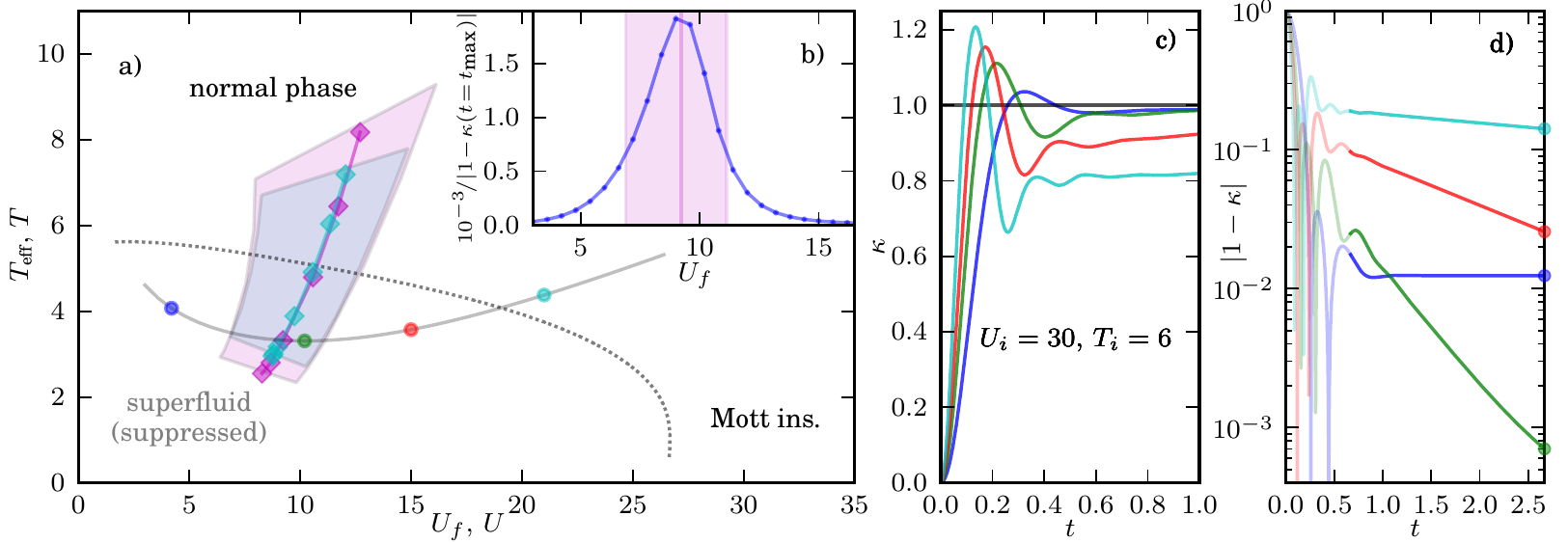}
\caption{\label{fig:UTPhaseDiagram} (color online)
a) Non-equilibrium $(U_f, \Teff)$ ``phase diagram'' for quenches within the symmetric Mott and normal phases ($|\phi| = 0$) with $\langle \n \rangle = 1$. While the superfluid state is absent, the equilibrium superfluid $(U, T)$ phase boundary is shown for guidance (dotted line).
The shaded areas indicate the occurrence of rapid thermalization, for the initial interactions $U_i=30$ (cyan area) and $45$ (magenta area) and $T_i \in [3, 18]$. The point of most rapid thermalization is defined as $U_f$ maximizing $|1-\kappa(t=\tmax)|^{-1}$ (diamonds), and the left and right boundaries of the rapid thermalization area correspond to a three-fold decrease from the maxima, as explicitly shown for $U_i=30$ and $T_i=6$ in (b).
For the quenches from $U_i=30$ and $T_i=6$ ending at $(U_f, \Teff)$ (solid gray line) the real-time evolution of $\kappa(t)$ for $U_f = 4.2$, $10.2$, $15$, and $21$ (blue, green, red, and cyan lines) (see circles in a)), are shown in c) for short times and in d) for long times with $|1-\kappa(t=\tmax)|$ (markers).
} \end{figure*}

BDMFT captures the conversion between interaction, kinetic, and condensate energy, as well as the relaxation to the predicted thermal values (Fig.~\ref{fig_energies}). Despite a nontrivial time-evolution of the individual components the total energy $E_{t}$ and the particle number $\langle \n \rangle$ is conserved to high accuracy by our 5th order solver (right panel).

We should note, however, that the NCA solution yields an approximate spectral function, as for the Fermi-Hubbard model \cite{Pruschke:1993aa}. To assess these errors it is useful to check the accuracy to which spectral sum rules (valid also in a nonequilibrium setting \cite{Freericks:2013oq}) are fulfilled. 
The moments $\mu^R_n(T)$ of the spectral function $A^R(T, \omega)$, $\mu^R_n(T) = \int_{-\infty}^\infty d\omega \, \omega^n A^R(T, \omega)$, are given by the higher order derivatives of the retarded Green's function $G^R(T, t)$ at $t=0^+$, $\mu^R_n(T) = - \textrm{Im} [ i^n \partial_t^n G^R(T, t) ]_{t=0^+}$, where $T$ and $t$ are the absolute and relative time respectively, see Ref.\ \onlinecite{Freericks:2013oq}. The moments can also be determined using the equation of motion, in terms of operator expectation values, $\mu^R_0 = 1$, $\mu^R_1 = \langle \epsilon \rangle -\mu + 2\langle \n \rangle U$, and $\mu^R_2 = \langle \epsilon^2 \rangle + \mu^2 + 3U^2\langle \n^2 \rangle - \langle \n \rangle(4\mu U + U^2)$, where $\langle \epsilon^n \rangle$ denotes the $n$th moment of the non-interacting density of states, see Ref.\ \onlinecite{Anders:2011uq}. For an approximate solution of the BDMFT equations, these approaches do not yield the same result.
In equilibrium, BDMFT+NCA gives a 1.6\% relative error of the first spectral moment $\mu^R_1$ in the Mott insulating phase ($U=96$, $T=6$), and a 10\% error in the vicinity of the superfluid phase boundary ($U=40$, $T=6$). For the second moment $\mu^R_2$, the relative errors are 11\% and 36\%, respectively. 

\subsection{Quenches from the Mott insulator} \label{sec:Mott}

As a first application, we study quenches within the symmetric Mott and normal phases ($|\phi|=0$), i.e.\ suppressing symmetry breaking (superfluid states). Since the Gutzwiller mean-field description only contains the condensate energy $E_c$ and the interaction energy $E_i$, the symmetric state in this approximation is simply the atomic limit with $E_c \propto |\phi|^2=0$. So,  for these quenches, no energy conversion occurs in the Gutzwiller treatment, resulting in an unphysical constant time-evolution. 
BDMFT, however, retains temporal fluctuations and enables the conversion of interaction energy $E_i$ into kinetic energy $E_k$, and vice versa, which leads to a non-trivial quench dynamics. 

To search for thermalization we study the relative change $\kappa$ in the double occupancy $\langle \n^2 \rangle$,
\begin{equation}
 \kappa = \frac{\langle \n^2 \rangle(t) - \langle \n^2 \rangle_{U_i, T_i}}
        {\langle \n^2 \rangle_{U_f, \Teff} - \langle \n^2 \rangle_{U_i, T_i}} 
  \label{eq:kappa} \, ,
\end{equation}
defined so that $\kappa(t=0) = 0$ and $\kappa = 1$ for a thermalized state.
Using this quantity, we identify an intermediate region of rapid thermalization in the $(U, T)$ plane [Fig.\ \ref{fig:UTPhaseDiagram}a] in the following way: For a given initial state $(U_i, T_i)$ we locate the maximum of $|1-\kappa(t=\tmax)|^{-1}$ as a function of $U_f$ at the longest accessible time $\tmax=2.66$, as shown explicitly for $U_i=30$ and $T_i=6$ in Fig.\ \ref{fig:UTPhaseDiagram}b. The values of $U_f$ and the corresponding effective temperatures $T_\text{eff}$ are shown in Fig.\ \ref{fig:UTPhaseDiagram}a, and the result turns out to be insensitive to the initial interaction ($U_i = 30$, $45$). In both the weak and strong-coupling $U_f$ regimes the system is trapped in a long-lived ``prethermalized state'',  reminiscent of the relaxation dynamics in the paramagnetic Fermi-Hubbard model \cite{Eckstein:2009fu}. 
%
The observed absence of thermalization in these regimes can be understood in terms of proximity to an integrable point, since the Bose-Hubbard model is integrable both for $U=0$ and $U=\infty$ \cite{Sorg:2014vn}.
%
Interestingly the relaxation behavior in the two regimes differ. In the strong-coupling regime the exponential decay of $|1 - \kappa|$ slows down as $U_f$ increases (green, red, cyan lines in Fig.\ \ref{fig:UTPhaseDiagram}d). In the low $U_f$ regime $\kappa$ very rapidly reaches a plateau value (blue line in Fig.\ \ref{fig:UTPhaseDiagram}d), which increases roughly exponentially as $U_f$ is decreased. 
The crossover between these two disparate behaviors is hard to pinpoint, and the indicated regions in Fig.\ \ref{fig:UTPhaseDiagram}a are only qualitative as $|1-\kappa(t=\tmax)|^{-1}$ is $\tmax$ dependent (the region becomes narrower and shifts to slightly higher $U_f$ with increasing $\tmax$).

\subsection{Quenches from the superfluid} \label{sec:SuperFluid}

\subsubsection{Relaxation regimes} \label{sec:SuperFluidRegimes}


\begin{figure}
\includegraphics[scale=1]{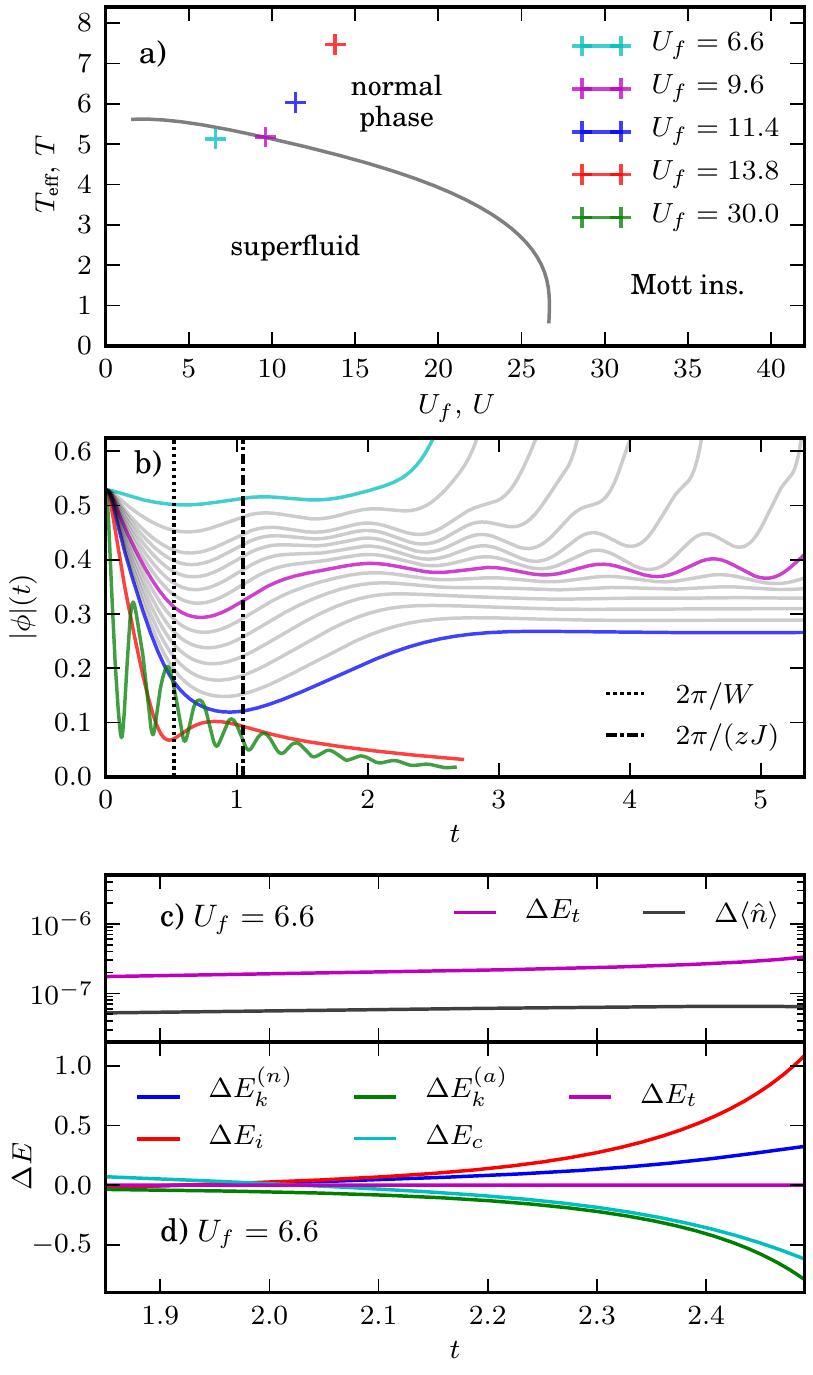}
\caption{\label{fig:SmallUf} (color online) Interaction quenches starting in the superfluid state ($U_i=6$, $T_i=5.1$), with
a) the positions of the final states ($U_f$, $\Teff$) (crosses) and the $(U, T)$ equilibrium superfluid phase boundary (gray line), and b) the corresponding evolution of the magnitude of the order parameter $|\phi|(t)$.
For $U_f=6.6$ the (equilibrium) thermal reference state at $(U_f, \Teff)$ is also superfluid and the non-equilibrium dynamics displays a rapid transient growth of the condensate (cyan). Close to the phase boundary in the normal phase the system is trapped in a superfluid for long times and the quench induced excitation is transferred to an amplitude mode at longer times (magenta). For intermediate $U_f$ a constant trapped superfluid persists (blue) and after passing the dynamical transition $\Ucdyn$ the system undergoes exponential relaxation to the normal phase (red). 
For large final interactions $U_f \gg W,zJ$ the condensate displays ``collapse-and-revival" oscillations with frequency $\omega \approx U_f$ (green). The evolution from $U_f=6.6$ to $U_f=11.4$ is also shown (gray lines).
For the growing condensate at $U_f=6.6$ panel c) shows the accuracy in total energy $E_t$ and density $\langle \n \rangle$ and panel d) shows the energy conversion during the time of rapid condensate growth.
} \end{figure}

The non-equilibrium dynamics after a quench from the superfluid ($|\phi| > 0$) with weak interaction $U_i = 6$ to larger interactions $U_f > U_i$ generates a variety of dynamical behaviors depending on $U_f$.
Apart from $U_f$ the system has two other characteristic energies (or inverse timescales), namely the bandwidth $W=12J$ and the condensate coupling $zJ = 6J$ (where $J=1$ is our unit of energy). In general the time evolution can be separated into five regimes, see Fig.\ \ref{fig:SmallUf}a and \ref{fig:SmallUf}b.

i) For quenches deep into the Mott regime, i.e.\ for large $U_f \gg W, zJ$, the condensate oscillates with the frequency $\omega$ of the final interaction strength, $\omega \approx U_f$, while relaxing exponentially (green line).
The relaxation rate strongly depends on the initial temperature $T_i$. For high $T_i$ (as in Fig.\ \ref{fig:SmallUf}b) the system displays relaxation to the Mott phase, while for low $T_i$ the system is trapped in a non-equilibrium superfluid state for long times, see Sec.\ \ref{sec:LongTimeDynamics}.

ii) In the intermediate coupling regime $U_f \gtrsim W > zJ$ the interaction driven oscillations compete with the kinetic time scale and only a few oscillations can be observed, and after the condensate time scale $2\pi/(zJ)$ the system displays exponential relaxation (red line).

iii) For $W \gtrsim U_f > zJ$ the thermal reference state is closer to the phase boundary in the normal phase. In this regime, after an initial transient undershoot in $|\phi|$, the system becomes trapped in a non-equilibrium superfluid state with a constant non-zero condensate (blue line).

iv) In the same range of $U_f$ an amplitude mode is excited at longer times (magenta line) with a roughly constant frequency but growing amplitude as the phase boundary is approached from the normal phase side.

v) For small $\Delta U = U_f - U_i$, ($W > U_f \approx zJ$) the initial transient is weak as the quench-energy scales with $\Delta U$. First the condensate undergoes a weak oscillatory transient followed by a sudden rapid growth (cyan line). This growth occurs when the final state is in the equilibrium superfluid region. 
The rapidly growing condensate is a numerical challenge because of the occupation of high-energy (i.e.\ high occupation-number) states.
On the one hand, the cutoff in the bosonic Fock space must be chosen large enough to accommodate this, and on the other hand, the fast oscillations of the high-energy modes require a small time-step.
In Fig.\ \ref{fig:SmallUf}b We plot the results up to the point to which they can be fully converged both in the size of the time-step and the size of the Hilbert space. For $N_\text{max}=11$ and $\Delta t = 0.005$ the drift in total energy $\Delta E_t = |E_t(t) - E_t(0)|$ and density $\Delta \langle \n \rangle = |\langle \n(t) \rangle - \langle \n(0) \rangle|$ is of the order $\lesssim 10^{-6}$, see Fig.\ \ref{fig:SmallUf}c.
Hence, we conclude that the growth is a robust feature of our BDMFT+NCA calculations.

During the growth there is a rapid conversion between the different energy components of the system, while the total energy $E_t$ is conserved, see Fig.\ \ref{fig:SmallUf}d. The interaction energy $E_i$ and the normal component of the kinetic energy $E_{k}^{(n)} \propto \langle b_i^\dagger b_{i+1} \rangle$ rapidly increase, while the condensate energy $E_c$ and the anomalous component of the kinetic energy $E_{k}^{(a)} \propto \langle b_i b_{i+1} \rangle$ decrease by the same total amount.

The self-amplified transient growth of the condensate fraction resembles the quantum turbulence driven dual cascade with non-equilibrium Bose-Einstein condensation observed in scalar field theories \cite{Berges:2012uq}.
Also in other contexts, dynamical instabilities with diverging solutions have been observed in lattice boson systems. In the weak coupling limit, a Gross-Pitaevskii treatment yields dynamically unstable solutions \cite{Machholm:2003ly}, also observed experimentally \cite{De-Sarlo:2005kl}. In the Bose-Hubbard model the exponential condensate growth of symmetric initial states has previously been studied in mean-field \cite{Snoek:2011hc}.

However, for the weak interaction quenches within the superfluid we cannot rule out that the sudden condensate growth is an artifact of the NCA treatment. Higher-order implementations of the self-consistent strong coupling expansion may in the future help to clarify this issue. 

\subsubsection{Short time dynamics 
after quenches deep into the Mott regime} \label{sec:ShortTimeDynamics}


\begin{figure}[b]
\includegraphics[scale=1]{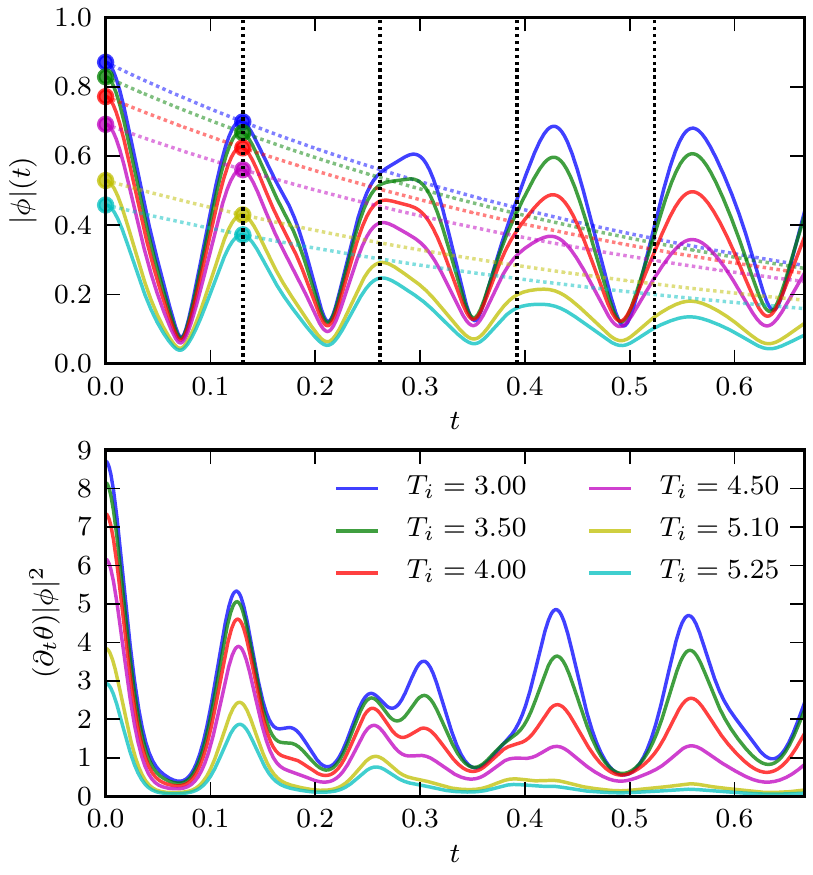}
\caption{\label{fig:TiSweep} (color online) Superfluid quench short time dynamics for $U_i=6$ and $U_f=48$ deep within the Mott phase ($U_f \gg W > zJ$). Upper panel: The first revival maximum coincides with the final interaction period $n \cdot 2\pi/U_f$ (black dotted lines). Exponential fits for the relaxation, using the first revival maximum at $t = 2\pi/U_f$ (colored dotted lines) are also shown. Lower panel: Time dependence of $(\partial_t \theta) |\phi|^2$.
} \end{figure}


\begin{figure*}
\includegraphics[scale=1]{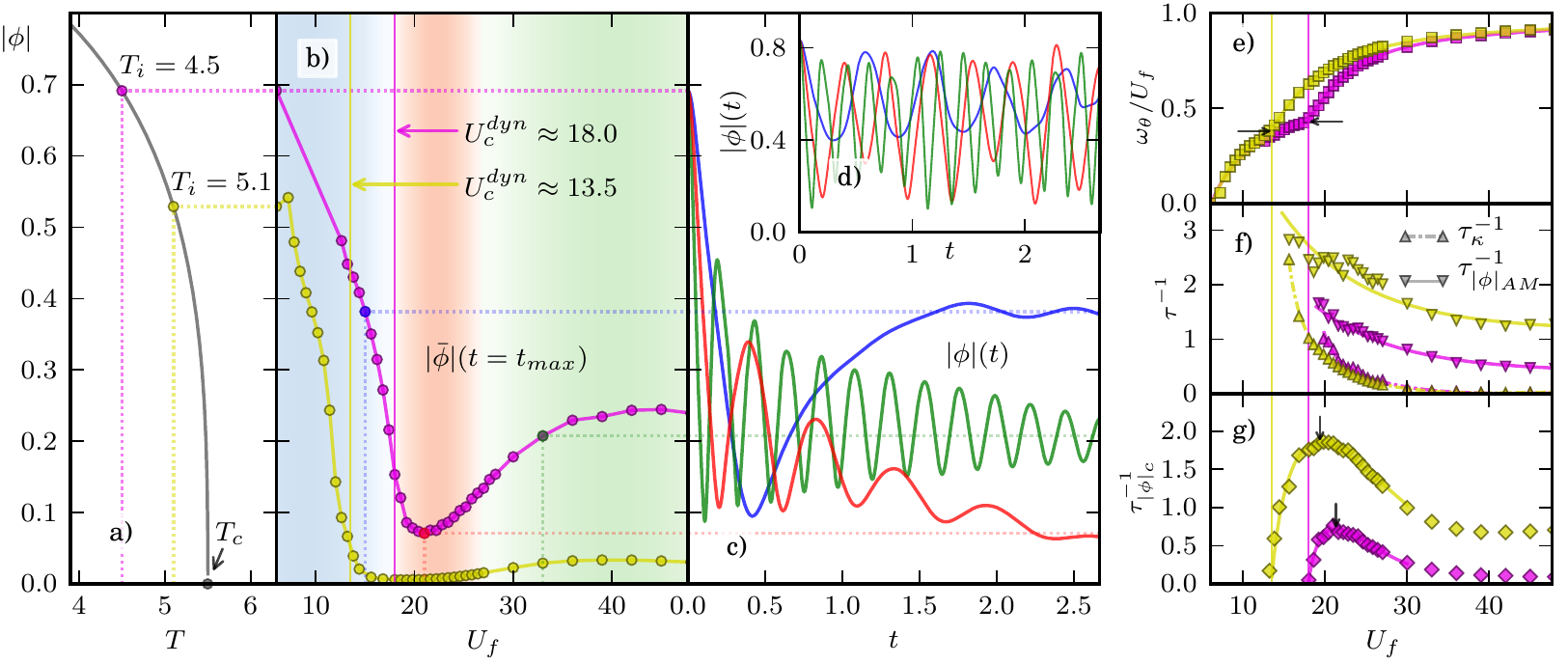}
\caption{\label{fig:SFquench} (color online) a) Interaction quenches from superfluid initial states with $U_i=6$, $\langle \n \rangle = 1$, and initial temperatures $T_i = 4.5$ (magenta) and $5.1$ (yellow), close to $T_c \approx 5.49$.
b) The window averages $\bar{|\phi|}(t=\tmax)$ are suppressed in the crossover region (red shaded region). c) Typical time evolutions of $|\phi|$ (solid lines) and $\bar{|\phi|}(t=\tmax)$ (dotted lines), for $T_i = 4.5$, in the three regimes and d) mean-field results for the same parameters are also shown.
e) The phase frequency $\omega_{\theta}$ exhibits a kink (arrows) at the dynamical transition $\Ucdyn \approx 13.5$ and $18.0$ where f) the double occupancy relaxation $\tau^{-1}_{\kappa}$ is peaking and the damping of the amplitude-mode $\tau^{-1}_\AM$ is maximal, while g) the condensate amplitude relaxation $\tau^{-1}_\CC$ peaks after $\Ucdyn$ (arrows).}
\end{figure*}

In the limit of large final interaction $U_f \gg W$, $zJ$, i.e.\ in regime (i), the superfluid quenches from $U_i = 6$ display oscillations with a frequency $\omega$ scaling with $U_f$, $\omega \approx U_f$, and the short time behavior is dominated by the interaction, as it defines the shortest time scale of the system.
The short time relaxation is expected to be driven by local decoherence and the long time relaxation ($t > 2\pi/W \approx 0.52$) to be dominated by hopping.
In order to study the short time dynamics we perform a series of quenches to $U_f = 48$ for several initial temperatures $T_i = 3.00, \dots , 5.25$, see Fig.~\ref{fig:TiSweep}.
While $\omega$ scales with $U_f$ there are important contributions from other frequency components. Pure $2\pi/U_f$-oscillations can only be observed in the first few revivals, while they are at later times washed out by the off-diagonal mixing of local occupation number states in the initial state.
The first revival maximum occurs at the period of the final interaction $2\pi / U_f$, the second revival has a pronounced two peak structure with the first peak occurring at $2 \cdot 2\pi/U_f$, and in the third revival the $3 \cdot 2\pi/U_f$ peak only appears as a shoulder of the main peak.
From Fig.\ \ref{fig:TiSweep} it is also evident that the short time decoherence strongly depends on the initial temperature $T_i$, with higher temperature resulting in faster damping.

An interesting question is whether the long time relaxation rate can already be inferred from the short time decoherence, in the spirit of the strong coupling analysis of Ref.\ \onlinecite{Fischer:2008ys}. To investigate this  we fit the simple exponential model $|\phi|(t) \approx |\phi|(0) e^{-t / \tau}$ to the real time data, where the relaxation rate $\tau$ is approximated using $|\phi|$ at the first revival maximum $t_r = 2\pi/U_f$ as $\tau = t_r / \log(|\phi|(t_r) /|\phi|(0))$.
Figure \ref{fig:TiSweep} shows that the relaxation rate $\tau$ is overestimated for low temperature initial states and underestimated for high temperature initial states. 
Hence in this regime the condensate relaxation can not be inferred from the first revival maximum.
Infact the long time exponential relaxation rate is only established after the characteristic condensate time scale $2\pi/(zJ)$, see e.g.\ the green and red lines in Fig.~\ref{fig:SmallUf}b.

We also note that the BDMFT calculation does not involve any approximation concerning the timescales of the dynamics. This sets it apart from, for example, the low frequency approximation applied in the Schwinger-Keldysh generalization of the strong coupling approach \cite{Kennett:2011fk}, where in the particle-hole symmetric limit $\langle \n \rangle = 1$ the condensate phase $\theta$ and amplitude $|\phi|$ (where $\phi = |\phi|e^{i\theta}$) are constrained by $(\partial_t \theta) |\phi|^2 = C$ for some constant $C$. As shown in the lower panel of Fig.~\ref{fig:TiSweep}, the BDMFT dynamics has a non-trivial time dependence in this quantity.

\subsubsection{Long time dynamics}
\label{sec:LongTimeDynamics}

The long time dynamics for quenches from the superfluid has been investigated in a number of zero temperature Gutzwiller mean-field studies \cite{Huber:2007ys, Wolf:2010bh, Sciolla:2010uq, Sciolla:2011kx, Snoek:2011hc, Krutitsky:2011qf}. The most prominent nonequilibrium effect is a dynamical transition at $U_f=\Ucdyn[U_i]$ \cite{Sciolla:2010uq}.
It is important to note, however, that for low $U_i$ only the mean-field calculation using a constrained basis including the lowest three bosonic occupation number states ($\Nmax= 3$) produces a sharp transition. If the physically important states with higher occupations are also considered, the transition turns into a crossover 
\footnote{Note that the $\Nmax > 3$ results of Refs.\ \onlinecite{Sciolla:2010uq} and \onlinecite{Sciolla:2011kx} are in the high-$U_i$ regime.}.
In a broader perspective the occurrence of a dynamical transition is not specific to the Bose-Hubbard model, but has also been observed in the Fermi-Hubbard model and other systems on the mean-field level \cite{Schiro:2010fk, Gambassi:2011dz, Sciolla:2011kx}. 

The quantum fluctuations missing in mean-field treatments are expected to heavily modify the dynamical transition, as previously shown for other systems, using dynamical mean field theory \cite{Eckstein:2009fu}, the Gutzwiller approximation including Gaussian fluctuations \cite{Schiro:2011oq, Sandri:2012fv}, and $1/N$ expansions \cite{Sciolla:2013bs}.
Here we show how the dynamical transition in the Bose-Hubbard model is affected when we go beyond the simple mean-field treatment, starting from a thermal initial state, and include quantum fluctuations using BDMFT.

As the hopping induced relaxation is most prominent for small $U_i$ and temperatures $T_i$ close to the superfluid phase boundary, we fix $U_i=6$, far away from the zero temperature transition, $U_c(T=0) \approx 26.4$, see Fig.~\ref{fig:PhaseDiag}. To see the enhanced relaxation in the vicinity of the phase boundary, located at $T_c(U=6) \approx 5.49$, we consider the two initial temperatures $T_i = 4.5$ and $5.1$ with relatively weak superfluidity $|\phi|^2 \lesssim 0.5$, see Fig.\ \ref{fig:SFquench}a.


To analyze the dynamics of the condensate amplitude $|\phi|$
we first look at windowed time averages $\bar{|\phi|}(t)$, using a Gaussian window with width $t_w = 1/3$ to filter out oscillations.
In Fig.\ \ref{fig:SFquench}b we plot the window average $\bar{|\phi|}(t=\tmax)$ at the longest time as a function of $U_f$, thereby restricting ourselves to the regimes (i)-(iii) above, i.e., when the final equilibrium state is not in the superfluid phase and the order parameter does not show self-amplified growth.
From Fig.\ \ref{fig:SFquench}b it is evident that $\bar{|\phi|}(t=\tmax)$ exhibits a crossover for $T_i=4.5$, from high values at low $U_f$ close to $U_i=6$, through a minimum at intermediate $U_f$, and increasing again for $U_f \gtrsim 30$, in qualitative agreement with the mean-field dynamical transition \cite{Sciolla:2010uq}. Also the general temperature dependence, namely that a higher temperature leads to lower condensate averages, agrees with mean-field.
However the thermal effects in BDMFT are much stronger: both the minimum and the large $U_f$ plateau are drastically reduced, going from $T_i=4.5$ to $T_i=5.1$. As we will show, this reduction is due to a rapid condensate relaxation rate $\tau^{-1}_\CC$ emerging close to the phase boundary (Fig. \ref{fig:SFquench}g).

The BDMFT real-time evolution in the three regimes is shown in Fig.\ \ref{fig:SFquench}c. For small $U_f$ in regime (iii), $|\phi|$ stabilizes at a finite value after an initial transient, even though the thermal reference state is in the normal phase. The intermediate regime (ii) shows fast thermalization with a rapidly decaying condensate and damped collapse-and-revival oscillations, in qualitative agreement with the experimental results of Greiner {\it et al.~}\cite{Greiner:2002fk}. Interestingly, for larger $U_f$ in regime (iii) the system is again trapped in a nonthermal superfluid phase, now exhibiting coherent amplitude oscillations with finite life-time.
Note that none of these relaxation and thermalization effects are captured by the Gutzwiller mean-field approximation, which predicts an oscillatory behavior (Fig.~\ref{fig:SFquench}d). 

While the minimum of $\bar{|\phi|}(t=\tmax)$ indicates a crossover, the dynamical transition $\Ucdyn$ can be accurately located by studying the condensate phase $\theta(t)$, where $\phi(t) = |\phi|(t) \, e^{i\theta(t)}$. 
Its linear component $\partial_t \theta(t) \approx \omega_{\theta}$ exhibits a kink at $U_f = \Ucdyn$, see Fig.\ \ref{fig:SFquench}e, in direct analogy with mean-field predictions where $\theta$ (mapped to a conjugate momentum $p$) also has a slope discontinuity at $\Ucdyn$ \cite{Sciolla:2010uq}.
Similar to the fast thermalization region in the symmetric phase, the double occupancy thermalizes rapidly for $U_f \approx (\Ucdyn)^+$ as can be seen in Fig.\ \ref{fig:SFquench}f from the drastic increase in the relaxation $\tau_{\kappa}^{-1}$ of 
\begin{equation*}
  |1-\kappa| \propto e^{-t / \tau_{\kappa}} \, ,
\end{equation*}
as $U_f \rightarrow (\Ucdyn)^+$.

To get a qualitative understanding of the condensate amplitude $|\phi|$ relaxation dynamics, we fit the late time-evolution $(t>1.3)$ to a damped two component model
\begin{equation*}
  |\phi_M|(t) = 
  A_\CC e^{-t/\tau_\CC} + A_\AM \cos^2(\omega t+\varphi) e^{-t/\tau_\AM}, 
\end{equation*}
with a non-oscillatory component ($\CC$) and a coherent amplitude-mode ($\AM$), and relaxation $\tau^{-1}_\CC$ and damping $\tau^{-1}_\AM$ respectively, see Fig.\ \ref{fig:SFquench}f and \ref{fig:SFquench}g. 
The amplitude-mode frequency $\omega$ has the same general behavior as $\omega_{\theta}$ (not shown), and $\omega, \omega_{\theta} \rightarrow U_f$ in the large $U_f$ limit.
Analogous to the rapid relaxation of the double occupancy, the amplitude mode is strongly damped for $U_f \approx (\Ucdyn)^+$, but it retains a finite lifetime $\tau^{-1}_\AM > 0$ for large $U_f$, see Fig.\ \ref{fig:SFquench}f. The relaxation of the non-oscillatory component shows two distinct behaviors: For $U_f < \Ucdyn$ the system is trapped in a superfluid state and the condensate relaxation is almost zero, $\tau_\CC^{-1} \approx 0$, while for $U_f \gtrsim \Ucdyn$ it becomes finite, reaching a maximum at intermediate $U_f$, see Fig.\ \ref{fig:SFquench}g.
For large $U_f$ and $T_i=5.1$, $\tau_\CC^{-1}$ stays finite and the system eventually thermalizes to the Mott state, while for $T_i=4.5$, $\tau_\CC^{-1}$ becomes small as $U_f \rightarrow \infty$, which means that the system is trapped for a very long time in a superfluid state. 
The stability of the superfluid can be understood in terms of a simple two fluid model of doublons and hard-core bosons \cite{Sorg:2014vn}. In this picture the quench generates long-lived doublons and depletes the hard-core boson gas away from unity filling, where it can remain a superfluid for any local interaction \cite{Schneider:2014ve}.
This case is particularly interesting as it opens up the possibility to study the Higgs amplitude mode in a metastable superfluid.

\subsubsection{Nonequilibrium ``phase diagram"}
\label{sec:SuperFluidPhaseDiagram}


\begin{figure}
\includegraphics[scale=1]{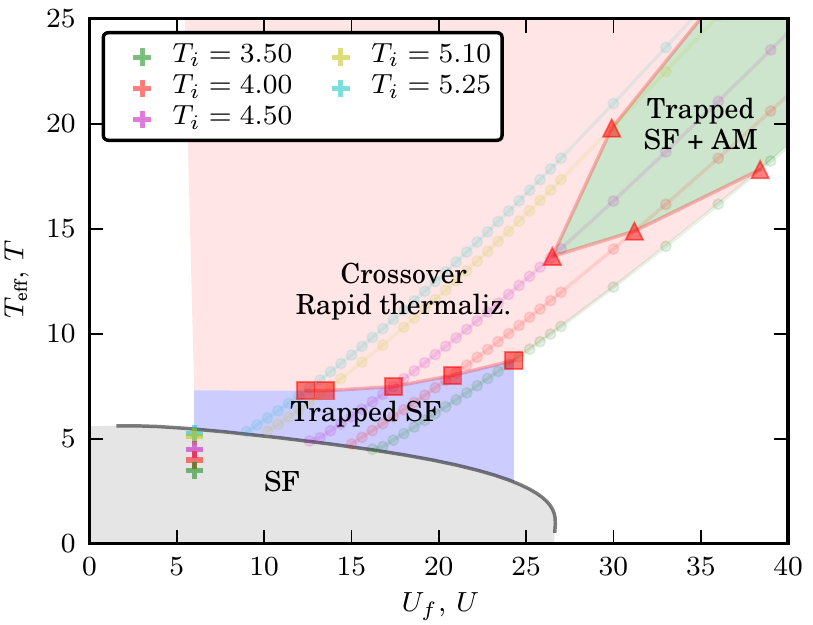}\caption{\label{fig:PhaseDiagram} (color online)
Qualitative non-equilibrium $(U_f, \Teff)$ phase diagram for quenches from equilibrium states (crosses) in the superfluid $(U, T)$ phase region (gray), with $U_i = 6$ and initial temperatures $T_i = 3.50$, $4.00$, $4.50$, $5.10$, and $5.25$, to final states with $(U_f, \Teff)$ (circles) within the equilibrium normal phase region.
The boundary between the trapped superfluid (SF) (blue) and the crossover region (red) is given by $\Ucdyn$ (squares), and the boundary between the crossover (red) and trapped superfluid with amplitude mode (SF+AM) (green) is taken as the point where $\tau^{-1}_\CC$ drops below $50\%$ of its crossover peak-value on the high-$U_f$ side (triangles), see Fig.~\ref{fig:SFquench}g.
} \end{figure}

We summarize the results for the long-time dynamics in the non-equilibrium ``phase diagram" shown in Fig.~\ref{fig:PhaseDiagram}. 
By repeating the analysis for $U_i=6$ and the series $T_i = 3.50$, $4.00$, $4.50$, $5.10$, and $5.25$ of initial temperatures, we locate the boundaries of the three dynamical regimes in the equilibrium normal phase region, regime (i), the high-$U$ region characterized by a trapped superfluid and amplitude mode (green), regime (ii), the crossover region with rapid thermalization (red), and regime (iii), the trapped superfluid in the vicinity of the equilibrium superfluid phase (blue).
While this non-equilibrium ``phase diagram'' depends on the initial states and the quench protocol, it gives an overview of the different relaxation and trapping phenomena and their location in parameter space.
An experimental verification of these different dynamical regimes of the Bose-Hubbard model would be very interesting and presumably possible. 

In the one dimensional Bose-Hubbard model a similar behavior has been theoretically observed using DMRG \cite{Kollath:2007ys}, and for longer times using time-dependent variational Monte Carlo \cite{Carleo:2012zr}. Quenches from the zero temperature superfluid display a region of thermalization at intermediate final interactions and a trapping in nonthermal states for long times at strong final coupling \cite{Kollath:2007ys}. At unity filling this behavior can be understood in terms of a reduced effective scattering of holon and doublon excitations at strong interactions \cite{Altman:2002ve}. Our results from BDMFT indicate that this phenomenon is also relevant in three dimensions (green region in Fig.\ \ref{fig:PhaseDiagram}). However, we also identified a transient trapping at low interactions (blue region in Fig.\ \ref{fig:PhaseDiagram}) which has not been reported for 1D. It is an open question whether this feature is specific to high-dimensional models. 
We also note that while BDMFT allows us to compare nonequilibrium and equilibrium states within the same formalism, the DMRG studies involve comparisons between time-dependent correlators and finite-temperature QMC results \cite{Kollath:2007ys}.

\section{Conclusions} \label{sec:Conclusions}

We developed the nonequilibrium BDMFT formalism and its implementation in combination with a NCA type bosonic impurity solver. We have demonstrated its ability to capture nontrivial dynamical effects in quenched Bose-Hubbard systems, including dynamical transitions, fast-thermalization crossovers, and trapped superfluid phases with long-lived but damped amplitude oscillations. These results were collected into two nonequilibrium ``phase diagrams'' (Figs.\ \ref{fig:UTPhaseDiagram} and \ref{fig:PhaseDiagram}), which illustrate the transitions and crossovers that occur as one varies the quench parameters. Particularly noteworthy results are the prediction of a very long-lived transient superfluid state with an amplitude mode after quenches from the superfluid phase into the Mott regime, our finding of a trapped superfluid state after quenches into the vicinity of the superfluid phase boundary, and the nonequilibrium Bose condensation (growing condensate) after small quenches within the superfluid phase.   

The ability of BDMFT to describe hopping induced relaxation phenomena at finite temperature goes beyond all current competing theoretical approaches. The Gutzwiller mean-field formalism lacks all hopping induced phenomena \cite{Sciolla:2010uq, Sciolla:2011kx}, and the strong coupling based real-time approach \cite{Kennett:2011fk} is limited to zero temperature and slow dynamics. The hopping perturbation expansion \cite{Trefzger:2011uq} looks promising but has so far only been applied at zero temperature. A comparative study with the finite temperature extension of this approach would be very interesting.

Extensions of the nonequilibrium BDMFT formalism to multi-flavor Hamiltonians \cite{Soyler:2009ys,Capogrosso-Sansone:2010zr} and inhomogeneous systems (e.g.\ with a trapping potential) \cite{Eckstein:2013fv} should enable direct comparisons with cold atom experiments.
Multi-orbital effects such as virtual excitations to higher orbitals can trivially be included in BDMFT in terms of effective three body interactions \cite{Johnson:2009kx}. 
While calculations based on unitary-time evolution \cite{Tiesinga:2011ve} suffice to understand experiments \cite{Will:2010uq} in the $J \rightarrow 0$ limit, BDMFT can extend the theoretical treatment to finite $J$.

An inhomogeneous extension of BDMFT will require a more advanced parallelization scheme than that applied by Dirks et al.\ \cite{Dirks:2014ij} for the fermionic case, but it would enable studies of very important phenomena, such as mass transport and the effects of the trapping potential in general cold-atom systems out-of-equilibrium.
The big challenge in these systems is the inherent disparity of the hopping and mass transport time-scales.
Simpler approximations such as the hopping expansion has successfully been applied in this context \cite{Dutta:2014uq},
but without incorporating thermal and retarded correlation effects.



In a broader perspective it should be productive to apply nonequilibrium BDMFT or variants of this formalism to nonequilibrium Bose-condensation in, e.g., polaritonic systems and field theories \cite{Carusotto:2013kx, Berges:2012uq, Akerlund:2013fk}. 

\begin{acknowledgments}
The authors would like to acknowledge fruitful discussions with 
H. Aoki, 
T. Ayral, 
J. Berges, 
N. Buchheim, 
D. Golez, 
F. Heidrich-Meisner, 
A. Herrmann, 
S. Hild, 
D. H\"{u}gel, 
M. P. Kennett,
Y. Murakami, 
L. Pollet, 
U. Schneider,  
N. Tsuji, 
and
L. Vidmar. 
The calculations have been performed on the UniFr cluster. HS and PW are supported by FP7/ERC starting grant No.\ 278023. 
\end{acknowledgments}

\appendix
\section{Nonequilibrium bosonic dynamical mean field theory}
\label{app:BDMFT}

The bosonic dynamical mean field theory (BDMFT) for the Bose-Hubbard model in \emph{equilibrium} has been derived in Ref.~\cite{Anders:2011uq} in three alternative ways, using the kinetic energy functional, an effective medium approach, and the quantum cavity method, which is very similar to the cavity construction by Snoek and Hofstetter \cite{Snoek:2010uq}. 
In this appendix we follow the latter approach, which combines the cavity construction with a generating functional formalism and a cumulant expansion to second order. By performing the derivation on the three branch Kadanoff-Baym contour $\C$ we obtain the \emph{nonequilibrium} generalization of BDMFT.
%
We will also show that BDMFT corresponds to the first order correction in the inverse coordination number $1/z$, and the second order correction in the fluctuations, of the mean-field approximation for the Bose-Hubbard model.
%

\subsection{The Bose-Hubbard model}

We consider the Bose-Hubbard model [Eq.\ (\ref{eq:Model})]
\begin{equation*}
  H = -J \sum_{\langle i, j \rangle} ( \bc_i\ba_j + \bc_j\ba_i )
 +\frac{U}{2} \sum_i \n_i(\n_i-1)  - \mu \sum_i \n_i \, ,
\end{equation*}
on a lattice with nearest neighbor hopping $J$ and a local pair interaction $U$, where $\n_i(\n_i-1) = \bc_i \bc_i \ba_i \ba_i$ is a pure two particle interaction counting the number of pairs on site $i$, and $\la i, j \ra$ denotes the sum over all nearest neighbor pairs $i$ and $j$.
Using the Nambu-spinor notation $\bbc = ( \bc, \, b )$ and collecting the local terms on site $i$ into $H_i = U \n_i(\n_i-1)/2 - \mu \n_i$, the Hamiltonian $H$ can be expressed as
\begin{equation}
  H = \sum_i H_i - J \sum_{\la i, j \ra} \bbc_i \bba_j \, ,
\end{equation}
where we have used that $\ba_i \bc_j = \bc_j \ba_i$ if $i \ne j$. Note that in the Nambu notation $\bbc_i \bba_j$ is hermitian, i.e.\
\begin{equation}
  \bbc_i \bba_j = \bbc_j \bba_i \, , \quad \textrm{for} \, i \ne j \, .
  \label{eq:NambuCommutation}
\end{equation}

\subsection{Kadanoff-Baym and Nambu formalism }

To treat an arbitrary time evolution starting from a finite temperature equilibrium state we formulate the theory on the three-branch Kadanoff-Baym contour $\mathcal{C}$ ($0 \! \rightarrow \! t_\text{max} \! \rightarrow \! 0 \! \rightarrow \! -i\beta$) \cite{Aoki:2014kx, Stefanucci:2013oq}. The partition function $\Z$ of the initial state can be expressed as $\Z = \Tr[ \TC e^{-iS} ]$ where $S$ is the action defined on the contour $\mathcal{C}$, $S = \int_\C dz \, H(z)$, $\mathcal{T}_\mathcal{C}$ is the time-ordering operator on $\mathcal{C}$, and the trace $\Tr[\cdot]$ runs over the Hilbert space of $H$. 
Time-dependent operator expectation values can be expressed in the trace formalism as
\begin{equation}
  \la \hat{O}(\ct) \ra_S = \frac{1}{\Z} \Tr[ \TC e^{-iS} \hat{O}(\ct) ] \, ,
\end{equation}
and the single-particle Green's function on the contour, $G(\ct, \ct')$, is given by $G(\ct, \ct') = -i \la \ba(\ct) \bc(\ct') \ra_S$. The Nambu generalization of the single-particle Green's function is a $2 \times 2$ matrix, which can be expressed in spinor notation as 
\begin{equation}
  \bG(\ct, \ct') = -i \la \Ba(\ct) \Bc(\ct') \ra \, . \label{eq:NambuGf}
\end{equation}
For a general introduction to the Kadanoff-Baym contour formalism, see Ref.\ \cite{Stefanucci:2013oq} and for a DMFT specific introduction see Ref.\ \cite{Aoki:2014kx}.

\subsection{Real-time generating functional}

To construct the generating functional on the contour $\C$ we introduce source fields $\eta_i$ on each site $i$ and the source action
\begin{equation}
  S_\eta = \int_\C d\ct \, H_\eta(\ct) \, , \quad \textrm{where }
  H_\eta = \sum_i \bbc_i \bea_i \, .
\end{equation}
Using $S_\eta$ and the action $S$ of the system
\begin{multline}
  S = \int_\C d\ct \sum_i H_i(\ct) - J \int_\C d\ct \sum_{\la i,j \ra} \bbc_i(\ct) \bba_j(\ct)
 \, ,
\end{multline}
the generating functional $\Z[\eta]$ can be defined as
\begin{equation}
  \Z[\eta] = \Tr \left[ \TC \exp[-iS + S_\eta] \right] \, .
\end{equation}
It can be used to compute any \emph{connected} response function by taking derivatives with respect to the source fields
\begin{equation}
  \frac{\partial^n}{\partial \ec_{\alpha_1} \dots \partial \ec_{\alpha_n}}
  \ln \Z[\eta] |_{\eta = 0} =
  \la \ba_{\alpha_n} \dots \ba_{\alpha_1} \ra_S^{(c)}
  \, .
  \label{eq:ZResponseFunction}
\end{equation}

\subsection{Cavity construction}

To derive a local effective action we use the standard cavity construction \cite{Georges:1996aa} and separate the Hamiltonian into three parts,
\begin{equation}
  H  = H_0 + \Delta H + H^{(0)} \, ,
\end{equation}
where $H_0$ acts on the site $i=0$, $\Delta H$ connects the zeroth site to its neighbors, and $H^{(0)}$ is the lattice with a cavity at the zeroth site, i.e.\
\begin{align}
  H_0 & = -\mu n_0 + \frac{U}{2} n_0 (n_0 - 1) \, , \\
  \Delta H & = - J \sum_{\la 0,i \ra} \bbc_i \bba_0 \, , \\
  H^{(0)} & = \sum_{i\ne 0} H_i 
    - J \sum_{\begin{subarray}{c} \la i,j \ra\\ i,j \ne 0\end{subarray}} 
      \bbc_i \bba_j \, ,
\end{align}
which in turn separates the action $S$ into
\begin{equation}
 S = S_0 + \Delta S + S^{(0)} \, .
\end{equation}
Analogously the source term can be decomposed into
\begin{equation}
  H_\eta = H_{0, \eta} + H^{(0)}_\eta , \,
\end{equation}
according to the same protocol, with
\begin{equation}
  H_{0, \eta} = \bbc_0 \bea_0 , \quad
  H^{(0)}_\eta = \sum_{i \ne 0} \bbc_i \bea_i \, ,
\end{equation}
which yields the corresponding terms of the source action 
\begin{equation}
 S_\eta = S_{0, \eta} + S^{(0)}_\eta \, .
\end{equation}

Using this separation of the zeroth site's degrees of freedom the generating functional can be written as
\begin{equation}
  \Z[\eta] = \Tr_0 \left[ \TC e^{-iS_0 + S_{0, \eta}}
  \Z^{(0)} \la e^{-i \Delta S + S^{(0)}_\eta} \ra_{S^{(0)}}
  \right] \, , \label{eq:GeneratingFunctional}
\end{equation}
where $\Tr_0[\cdot]$ denotes the trace over the Fock-space of the zeroth site.
In this form the generating functional can be approximated and/or taken to e.g.\ the infinite connectivity limit, which results in different types of dynamical mean field theory (DMFT) approximations.

\subsection{Cumulant expansion}

We are now ready to perform a \emph{cumulant expansion} \cite{JPSJ.17.1100} of the expectation value $\la e^{-i \Delta S + S^{(0)}_\eta} \ra_{S^{(0)}}$ in Eq.\ (\ref{eq:GeneratingFunctional}). Formally this corresponds to expanding $\ln \la e^{-i \Delta S + S^{(0)}_\eta} \ra_{S^{(0)}}$ in an infinite sum of response functions with respect to $S^{(0)}$. The initial logarithm ensures that the series enters in the exponent, and for this reason the procedure is often referred to as ``re-exponentiation''.

Following Ref.\ \onlinecite{JPSJ.17.1100} the cumulant expansion becomes
\begin{multline*}
  \ln \la \exp[-i \Delta S + S^{(0)}_\eta] \ra_{S^{(0)}} =
  \la \exp[-i \Delta S + S^{(0)}_\eta] - 1 \ra_{S^{(0)}}^{(c)} 
  \\ = \!
  \sum_{n=1}^\infty \frac{1}{n!} 
  \! \int_\C \!\! d\ct_1 \dots \!\! \int_\C \!\! d\ct_n
  \left\la \prod_{k=1}^n 
  ( -i \Delta H(\ct_k) + H^{(0)}_\eta (\ct_k) )
  \right\ra^{\!\! (c)}_{\!\! S^{(0)}}
  \! \! \! \! . \label{eq:CumulantExpansion}
\end{multline*}

In the derivation of the fermionic dynamical mean field effective action, the cumulant expansion terminates at second order in the limit of  infinite dimensions $z \rightarrow \infty$ (using a $J \rightarrow J/\sqrt{z}$ scaling of the hopping) \cite{Georges:1996aa}. This yields the usual hybridization function term
\begin{equation*}
  \ln \la e^{-i \Delta S + S^{(0)}_\eta} \ra_{S^{(0)}} = \dots =
  \iint_\C d\ct \, d\ct' \, \bc(\ct) \Delta(\ct, \ct') b(\ct') \, .
\end{equation*}

For Bosons, however, anomalous contributions due to symmetry breaking scale linearly with the coordination number $z$, requiring a $1/z$ scaling of the hopping to obtain a finite $z \rightarrow \infty$ limit \cite{Anders:2010uq, Anders:2011uq}. This procedure results in the mean field effective action \cite{Sachdev:1999fk} which does not include quantum fluctuations of non-condensed Bosons.
In order to retain fluctuations we therefore avoid taking the infinite connectivity limit and instead {\it truncate} the cumulant expansion at second order, which (as we will see) yields $1/z$ corrections in the effective action \cite{Snoek:2010uq}.

We write the second order approximation of the cumulant expansion as
\begin{equation}
  \ln \la \exp[-i \Delta S + S^{(0)}_\eta] \ra_{S^{(0)}}
  \approx
  -i S^{(0)}_{\textrm{eff}} + S^{(0)}_{\textrm{eff}, \eta} 
  \, ,
\end{equation}
collecting the source-free terms in the effective action 
\begin{multline}
  S^{(0)}_{\textrm{eff}} = 
  \int_\C \! d\ct \, \la \Delta H(\ct) \ra^{(c)}_{S^{(0)}}
  \\
  + \frac{i}{2} \iint_\C \! d\ct \, d\ct' \,
  \la \Delta H(\ct) \Delta H(\ct') \ra^{(c)}_{S^{(0)}}
  \, , \label{eq:S0effCumulant}
\end{multline}
and the terms containing source fields $\eta$ in the effective \emph{source} action 
\begin{multline}
  S^{(0)}_{\textrm{eff}, \eta} =
  \int_\C d\ct \, \la H^{(0)}_\eta(\ct) \ra^{(c)}_{S^{(0)}}
  - \frac{i}{2} \iint_\C d\ct \, d\ct' \,
  \\
  \times \Big[
  \la \Delta H(\ct) H^{(0)}_\eta (\ct') \ra^{(c)}_{S^{(0)}}
  + \la H^{(0)}_\eta (\ct) \Delta H(\ct') \ra^{(c)}_{S^{(0)}}
  \Big] 
  \\
  + \frac{1}{2} \iint_\C d\ct \, d\ct' \,
  \la H^{(0)}_\eta(\ct) H^{(0)}_\eta(\ct') \ra^{(c)}_{S^{(0)}}
  \, . \label{eq:S0effetaCumulant}
\end{multline}

Hence, by truncating the expansion at second order we obtain an effective action $S_{\textrm{eff}}$ and generating functional $\Z_{\textrm{eff}}[\eta]$ according to
\begin{multline}
  \frac{\Z[\eta]}{\Z^{(0)}} = 
  \Tr_0 \left[ \TC 
    e^{-iS_0 + S_{0, \eta}}
    \la e^{-i \Delta S + S^{(0)}_\eta} \ra_{S^{(0)}}
    \right]
  \\ \approx
  \Tr_0 \left[ \TC 
    \exp[-iS_0 - iS^{(0)}_{\textrm{eff}} + S_{0, \eta} + S^{(0)}_{\textrm{eff}, \eta}]
    \right]
  \\
  \equiv 
  \Tr_0 \left[ \TC 
    \exp[-iS_{\textrm{eff}}] \right]
  = \Z_{\textrm{eff}}[\eta] \, .
\end{multline}

\subsection{Explicit 2nd order form}

To obtain the explicit form for the local effective action 
\begin{equation}
 S_{\textrm{eff}} = S_0 + S^{(0)}_{\textrm{eff}} + i S_{0,\eta} + i S^{(0)}_{\textrm{eff},\eta}
 \label{eq:SeffComponents}
\end{equation}
we need to look into the details of the expansion giving the actions $S^{(0)}_{\textrm{eff}}$ and $S^{(0)}_{\textrm{eff},\eta}$. At a later stage, we will also make use of the effective generating functional in order to arrive at the contour generalization of the (self-consistent) B-DMFT effective action, previously derived for equilibrium in \cite{Anders:2011uq, Anders:2011fk}.

The operators appearing in the expansion of $S^{(0)}_{\textrm{eff}}$ and $S^{(0)}_{\textrm{eff},\eta}$ [Eqs.\ (\ref{eq:S0effCumulant}) and (\ref{eq:S0effetaCumulant})] are
\begin{align}
  \Delta H & 
  = -J\sum_{\la 0, i \ra} \bbc_i \bba_0 
  = -J\sum_{\la 0, i \ra} \bbc_0 \bba_i \, , \label{eq:DeltaH}\\
  H^{(0)}_\eta & 
  = \sum_{i \ne 0} \bbc_i \bea_i 
  = \sum_{i \ne 0} \bec_i \bba_i \, , \label{eq:H0eta}
\end{align}
where in the last steps we have used the hermitian property of Nambu creation-annihilation operator products [Eq.\ (\ref{eq:NambuCommutation})].
Hence the first order expectation values take the form
\begin{align}
  \la \Delta H(\ct) \ra^{(c)}_{S^{(0)}} & 
  = -J\sum_{\la 0, i \ra} \la \bbc_i(\ct) \ra^{(c)}_{S^{(0)}} \bba_0(\ct) \, ,\\
  \la H^{(0)}_\eta(\ct) \ra^{(c)}_{S^{(0)}} & 
  = \sum_{i \ne 0} \la \bbc_i(\ct) \ra^{(c)}_{S^{(0)}} \bea_i(\ct) \, .
\end{align}
The second order terms can be obtained using the two different ways of expressing the operators in Eqs.\ (\ref{eq:DeltaH}) and (\ref{eq:H0eta}) in order to arrive at Nambu response function expressions as in Eq.\ (\ref{eq:NambuGf}).
The second order term of $S^{(0)}_{\textrm{eff}}$ in Eq.\ (\ref{eq:S0effCumulant}) reads
\begin{multline}
  \la \Delta H(\ct) \Delta H(\ct') \ra^{(c)}_{S^{(0)}} 
  \\ =
  \bbc_0(\ct) 
  \Big[ \sum_{ \la 0,i \ra, \, \la 0, j \ra }
  J \la \bba_i(\ct) \bbc_j(\ct') \ra^{(c)}_{S^{(0)}} J
  \Big] \bba_0(\ct')
  \\ =
  i \bbc_0(\ct) 
  \Big[ \sum_{ \la 0,i \ra, \, \la 0, j \ra }
  J \bG^{(0)}_{ij}(\ct, \ct') J
  \Big] \bba_0(\ct')
  \\ = 
  i \bbc_0(\ct) 
  \bDelta(\ct, \ct') \bba_0(\ct') 
  \, ,
\end{multline}
 where we have introduced the connected single particle Green's function $\bG^{(0)}_{ij}(\ct, \ct') = -i \la \bba_i(\ct) \bbc_j(\ct') \ra^{(c)}_{S^{(0)}}$ of the lattice with cavity and the total hybridization function $\bDelta$ of the zeroth lattice site
\begin{equation}
  \bDelta(\ct, \ct') =
  \sum_{ \la 0,i \ra, \, \la 0, j \ra }
  J \bG^{(0)}_{ij}(\ct, \ct') J
  \, . \label{eq:bDeltaDefInG0}
\end{equation}
Hence, the source-free action $S^{(0)}_{\textrm{eff}}$ can be written as
\begin{multline}
  S^{(0)}_{\textrm{eff}} = 
  -J \int_\C d\ct \sum_{\la 0,i \ra} \la \bbc_i(\ct) \ra^{(c)}_{S^{(0)}} \bba_0(\ct)
  \\
  + \frac{1}{2} \iint_\C d\ct \, d\ct' \,
  \bbc_0(\ct) \bDelta(\ct, \ct') \bba_0(\ct')
  \, . \label{eq:S0eff}
\end{multline}

\subsubsection{Local effective source action}

Next we consider the second order terms of the effective source action $S^{(0)}_{\textrm{eff},\eta}$ [Eq.\ (\ref{eq:S0effetaCumulant})]. In terms of $\bG^{(0)}_{ij}(\ct, \ct')$ the quadratic source term reads
\begin{multline}
  \la H^{(0)}_\eta(\ct) H^{(0)}_\eta(\ct') \ra^{(c)}_{S^{(0)}} 
  \\ =
  i \sum_{i \ne 0 , \, j \ne 0}
  \bec_i(\ct) \bG^{(0)}_{ij}(\ct, \ct') \bea_j(\ct') \, ,
\end{multline}
and the first mixed term becomes
\begin{multline}
  \la \Delta H(\ct) H^{(0)}_\eta(\ct') \ra^{(c)}_{S^{(0)}} 
  \\ =
  -i \bbc_0(\ct)
  \sum_{\la 0, i \ra , \, j \ne 0} J \bG^{(0)}_{ij}(\ct, \ct') \bea_j(\ct')
  \, .
\end{multline}
By an interchange of integration variables it is possible to show that the other mixed term gives an equal contribution.

Collecting all the terms we arrive at the final expression for the local effective source action
\begin{multline}
  S^{(0)}_{\textrm{eff}, \eta} =
  \int_\C d\ct \sum_{i \ne 0} \la \bbc_i(\ct) \ra^{(c)}_{S^{(0)}} \bea_i(\ct)
  \\
  + \frac{i}{2} \iint_\C d\ct \, d\ct' \,
  \sum_{i \ne 0 , \, j \ne 0}
  \bec_i(\ct) \bG^{(0)}_{ij}(\ct, \ct') \bea_j(\ct')
  \\
  - \iint_\C d\ct \, d\ct' \,
  \bbc_0(\ct)
  \sum_{\la 0, i \ra , \, j \ne 0} J \bG^{(0)}_{ij}(\ct, \ct') \bea_j(\ct')
  \, ,
  \label{eq:S0effEta}
\end{multline}

\subsection{Local anomalous term}

We see in Eq.\ (\ref{eq:S0eff}) that the symmetry breaking of the infinite lattice system induces a local symmetry breaking term on the zeroth lattice site. The strength of the symmetry breaking field is however determined by the anomalous expectation values $\la \bbc_i(z) \ra^{(c)}_{S^{(0)}}$ on all sites $i$ \emph{neighboring the cavity}.

For finite coordination numbers $z$ the removal of the cavity site affects the neighboring sites, hence this expectation value is \emph{not} equal to that of the original homogeneous system \cite{Snoek:2010uq}
\begin{equation}
  \la \bbc_i(\ct) \ra^{(c)}_{S^{(0)}} 
  \ne \la \bbc_i(\ct) \ra^{(c)}_{S} 
  \approx \la \bbc_i(\ct) \ra^{(c)}_{S_{\textrm{eff}}} \, .
\end{equation}

To determine the difference between $\la \bbc_i(\ct) \ra^{(c)}_{S^{(0)}}$ and $\la \bbc_i(\ct) \ra^{(c)}_{S_{\textrm{eff}}}$ we calculate the latter using the effective generating functional $\Z_{\textrm{eff}}$ and the Nambu generalization of Eq.\ (\ref{eq:ZResponseFunction})
\begin{multline}
  \la \bbc_i(\ct) \ra^{(c)}_{S_{\textrm{eff}}}
  = \frac{\partial}{\partial \bea_i(\ct)} \ln \Z_{\textrm{eff}}[\eta] |_{\eta = 0}
  = \la \bbc_i(\ct) \ra^{(c)}_{S^{(0)}}
  \\ - \int_\C d\ct' \,
  \la \bbc_0(\ct') \ra^{(c)}_{S_{\textrm{eff}}}
  \sum_{\la 0, j \ra} J \bG^{(0)}_{ji}(\ct', \ct)
  \, .
\end{multline}
Thus the local anomalous term of $S^{(0)}_{\textrm{eff}}$ in Eq.\ (\ref{eq:S0eff}) can be rewritten, using only expectation values with respect to $S_{\textrm{eff}}$, as
\begin{multline}
  -J \int_\C d\ct \sum_{\la 0,i \ra} \la \bbc_i(\ct) \ra^{(c)}_{S^{(0)}} \bba_0(\ct)
  =
  -J \int_\C d\ct \sum_{\la 0,i \ra} 
  \Big[
  \la \bbc_i(\ct) \ra^{(c)}_{S_{\textrm{eff}}}
  \\ +
  \int_\C d\ct' \,
  \la \bbc_0(\ct') \ra^{(c)}_{S_{\textrm{eff}}}
  \sum_{\la 0, j \ra} J \bG^{(0)}_{ji}(\ct', \ct)
  \Big]
  \bba_0(\ct)
  \\ \shoveleft =
  \int_\C d\ct 
  \Big[
  -zJ \bPhi^\dagger_0(\ct)
  - \int_\C d\ct' \,
  \bPhi^\dagger_0(\ct')\bDelta(\ct', \ct)
  \Big] \bba_0(\ct) \, ,
  \label{eq:EffLocalAnomalousTerm}
\end{multline}
where in the last step, we have introduced the local anomalous amplitude $\bPhi_0^\dagger(\ct) = \la \bbc_0(\ct) \ra^{(c)}_{S_{\textrm{eff}}}$ and assumed translational invariance $\la \bbc_0(\ct) \ra^{(c)}_{S_{\textrm{eff}}} = \la \bbc_i(\ct) \ra^{(c)}_{S_{\textrm{eff}}}$.

\subsection{Local effective action}

Substituting Eqs.\ (\ref{eq:S0eff}) and (\ref{eq:S0effEta}) into Eq.\ (\ref{eq:SeffComponents}), rewriting the local symmetry breaking using Eq.\ (\ref{eq:EffLocalAnomalousTerm}) and setting the sources to zero $\eta = 0$, we obtain the BDMFT local effective action for the Bose-Hubbard model
\begin{multline}
  S_{\textrm{eff}} =
  \int_\C d\ct \left(
  -\mu n(\ct)
  + \frac{U}{2} n(\ct) ( n(\ct) - 1 )
  \right)
  \\ +
  \int_\C d\ct
  \Big[
  -zJ \bPhi^\dagger(\ct)
  - \int_\C d\ct' \,
  \bPhi^\dagger(\ct')\bDelta(\ct', \ct)
  \Big] \mbf{b}(\ct) 
  \\ + \frac{1}{2} \iint_\C d\ct \, d\ct' \,
  \bbc(\ct) \bDelta(\ct, \ct') \mbf{b}(\ct') \, ,
  \label{eq:EffectiveActionWithPhiEff}
\end{multline}
where we have dropped site indices and the complex field  $\bPhi^\dagger(\ct)$ is self-consistently defined as $\bPhi^\dagger(\ct) = \la \bbc(\ct) \ra_{S_{\textrm{eff}}}$.

\subsubsection{Equilibrium form}

On 
the imaginary time branch the field is constant $\bPhi^\dagger(\tau) = \bPhi^\dagger$, the hybridization function is time translational invariant $\bDelta(\tau, \tau') = \bDelta(\tau - \tau')$, and the action simplifies to 
\begin{multline}
  S_{\textrm{eff}} =
  \int_0^\beta d\tau \left(
  - \mu n(\tau)
  + \frac{U}{2} n(\tau) ( n(\tau) - 1 )
  \right)
  \\ +
  \bPhi^\dagger
  \Big[
  -zJ 
  -\int_0^\beta d\bar{\tau} \,
  \bDelta(\bar{\tau})
  \Big]
  \int_0^\beta d\tau
  \mbf{b}(\tau) 
  \\ + \frac{1}{2} \iint_0^\beta d\tau \, d\tau' \, 
  \bbc(\tau) \bDelta(\tau - \tau') \mbf{b}(\tau') \, ,
\end{multline}
in agreement with Refs.\ \onlinecite{Anders:2011uq, Anders:2011fk}, up to a minus sign on the hybridization function due to a different notation.

\subsection{One-loop correction in $1/z$}

To see that the BDMFT effective action in Eq.\ (\ref{eq:EffectiveActionWithPhiEff}) is a one-loop correction in the inverse coordination number $1/z$ one must study the scaling of its terms. %
The only non-trivial contribution comes from the hybridization function $\bDelta = J^2 \sum_{ \la 0,i \ra, \, \la 0, j \ra } \bG^{(0)}_{ij}$ [Eq.\ (\ref{eq:bDeltaDefInG0})].
On a graph without loops, such as the Bethe graph, the sum over nearest neighbors contains no cross-terms $\bG^{(0)}_{ij} = \delta_{ij} \bG^{(0)}_{ii}$, and the sum simplifies to
\begin{equation}
  \bDelta(\ct, \ct') = z J^2 \bG^{(0)}_{ii}(\ct, \ct') \, .
\end{equation}
On more general lattices, the power counting in $z$ gives the same leading order result, but is more elaborate \cite{Georges:1996aa}. 

Substituting this result into Eq.\ (\ref{eq:EffectiveActionWithPhiEff}) and making a $J \rightarrow J/z$ rescaling of the hopping gives the rescaled action
\begin{multline}
  \tilde{S}_{\textrm{eff}} =
  \int_\C d\ct \left(
  -\mu n(\ct)
  + \frac{U}{2} n(\ct) ( n(\ct) - 1 )
  \right)
  \\ +
  \int_\C d\ct
  \Big[
  -J \bPhi^\dagger(\ct)
  - \frac{J^2}{z} \int_\C d\ct' \,
  \bPhi^\dagger(\ct') \bG^{(0)}_{ii}(\ct', \ct)
  \Big] \mbf{b}(\ct) 
  \\ + \frac{J^2}{2z} \iint_\C d\ct \, d\ct' \,
  \bbc(\ct) \bG^{(0)}_{ii}(\ct, \ct') \mbf{b}(\ct') \, ,
\end{multline}
which makes it evident that the terms containing $\bG^{(0)}_{ii}$, i.e.\ the hybridization terms in the BDMFT effective action [Eq.\ (\ref{eq:EffectiveActionWithPhiEff})] corresponds to a $1/z$ correction of the mean-field action (obtained by setting $\bDelta = 0$).

\subsection{Second order fluctuation expansion}

While BDMFT can bee seen as a one-loop expansion in the inverse coordination number it is also a second order expansion in the condensate fluctuations, as discussed in Ref.\ \onlinecite{Anders:2011uq}. This can be made explicit by rewriting the terms containing the hybridization in the effective action using the fluctuation operators $\dBa$, defined as $\dBa \equiv \Ba - \Pa$ with $\langle \dBa \rangle = \mbf{0}$. Inserting these in Eq.~(\ref{eq:EffectiveActionWithPhiEff})  yields
\begin{multline}
  S_{\textrm{eff}} =
  \int_\C d\ct \left(
  -\mu n(\ct)
  + \frac{U}{2} n(\ct) ( n(\ct) - 1 ) - zJ \Pc(\ct) \Ba(\ct)
  \right)
  \\
  + \frac{1}{2} \iint_\C d\ct \, d\ct' \,
  \dBc(\ct) \bDelta(\ct, \ct') \dBa(\ct') \, ,
\end{multline}
where the hybridization term is the exact 2nd order contribution of the fluctuations. 
Hence BDMFT correctly describes the deep superfluid where fluctuations are suppressed, i.e.\ the weakly interacting Bose gas limit (WIBG) \cite{Capogrosso-Sansone:2010vn}.


\section{Nambu generalization of the non-crossing approximation}
\label{app:NCA}

The solution of impurity actions without symmetry breaking by means of self-consistent strong-coupling perturbation theory, i.e.\ the non-crossing approximation (NCA) and its higher-order generalizations, has been discussed in detail in Ref.\ \onlinecite{Eckstein:2010fk}.
To apply this method to the BDMFT action in Eq.\ (\ref{eq:Seff}) we have to extend the NCA formalism to Nambu spinors and symmetry broken states. 
The diagrammatics of Ref.\ \onlinecite{Eckstein:2010fk} needs to be modified on the operator and hybridization function level. 
While the pseudo-particle propagators $\hat{G}_{\Gamma \Gamma'}(\ct, \ct')$ still only carry local many-body state indices $\Gamma$ and $\Gamma'$ (corresponding to the occupation number states $| \Gamma \ra$ and $| \Gamma' \ra$), the hybridization function $\bDelta_{\gamma \nu}(\ct, \ct')$ now carries two Nambu indices $\gamma$ and $\nu$. 
We will represent the propagators with directed solid and dashed lines according to
\begin{equation*}
  \includegraphics{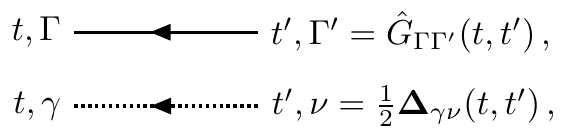}
\end{equation*}
where $\ct$ and $\ct'$ are times on the contour $\C$.

Due to the Nambu indices of $\bDelta$, the vertices of the theory must also be equipped with a Nambu index $\gamma$, in combination with a contour time $t$ and in and out going many-body state indices $\Gamma'$ and $\Gamma$, respectively. The matrix elements can be graphically represented as
\begin{equation*}
  \includegraphics{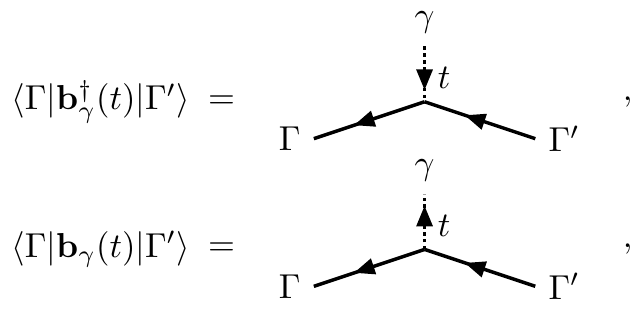}
\end{equation*}
where the direction of the hybridization line determines the operator of the vertex. A hybridization line entering a vertex creates a ``Nambu-particle'' by insertion of $\Bc_\gamma$ giving the matrix element $\la \Gamma | \Bc_\gamma | \Gamma' \ra$ , and an interaction line leaving the vertex annihilates a ``Nambu-particle'' through $\Ba_\gamma$ giving $\la \Gamma | \Ba_\gamma | \Gamma' \ra$. In the following we will use the operator symbols $\Bc_\gamma$ and $\Ba_\gamma$ to represent these matrix elements as they act in the same Fock-space as the pseudo-particle propagator $\Gp$ and the pseudo-particle self-energy $\Sp$.

\subsection{Pseudo particle self-energy}
\label{app:NCASigma}

Following the diagram rules of Ref.\ \onlinecite{Eckstein:2010fk} the pseudo-particle self-energy $\Sp$ at first order in $\bDelta$, corresponding to the non-crossing approximation (NCA), takes the form of shell-diagrams
\begin{equation*}
  \includegraphics{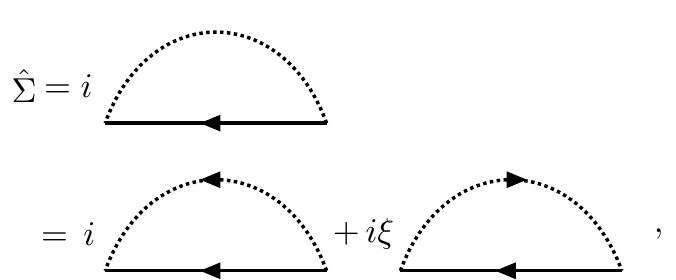}
\end{equation*}
where $\xi = \pm 1$ for bosons and fermions respectively.
Using the Nambu generalization of propagators and vertices gives the contour expression for the first diagram with a forward propagating hybridization line according to
\begin{multline} \\[-8mm]
  \hat{\Sigma}^{(1)}(\ct, \ct') = (i) 
  \includegraphics[valign=c]
    {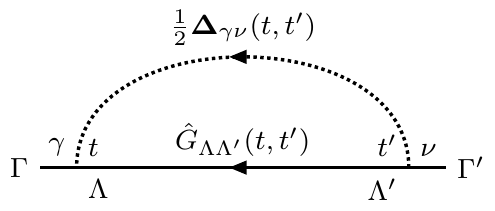}
  \\ =
  \frac{i}{2} \sum_{\gamma \nu } \bDelta_{\gamma \nu}(\ct, \ct')
  \big[ \mbf{b}^\dagger_\gamma(\ct) \, \hat{G}(\ct, \ct') \, \mbf{b}_\nu(\ct') \big]
  \, , \\[-8mm]
  \label{eq:NCASigma1}
\end{multline}
with implicit matrix multiplications in the many-body state indices $\Lambda$ and $\Lambda'$. The second diagram is constructed analogously
\begin{multline} \\[-8mm]
  \hat{\Sigma}^{(2)}(\ct, \ct') =
  (i\xi) 
  \includegraphics[valign=c]
    {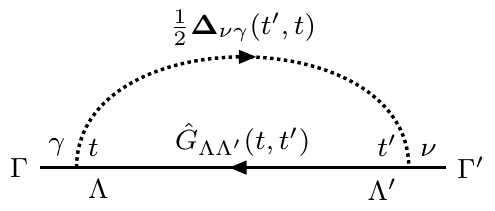}
  \\ =
  \xi \frac{i}{2} \sum_{\gamma \nu} \bDelta_{\nu \gamma}(\ct', \ct)
  \big[ \mbf{b}_\gamma (\ct) \, \hat{G}(\ct, \ct') \, \mbf{b}^\dagger_\nu(\ct') \big]
  \, . \\[-8mm]
  \label{eq:NCASigma2}
\end{multline}
Suppressing many-body state indices in these diagrams yields Fig.\ \ref{fig:NCA}a in Sec.\ \ref{sec:NCA}. Collecting all terms, we obtain
\begin{align*}
  \hat{\Sigma}(\ct, \ct') =
  & \frac{i}{2} \sum_{\gamma \nu}\bDelta_{\gamma \nu}(\ct, \ct') 
  \big[ \bbc_\gamma(\ct) \, \hat{G}(\ct, \ct') \,\bba_\nu(\ct') \big]
  \\ + 
  \xi & \frac{i}{2}\sum_{\gamma \nu}\bDelta_{\nu \gamma}(\ct', \ct)
  \big[ \bba_\gamma(\ct) \, \hat{G}(\ct,\ct') \,\bbc_\nu(\ct') \big]
  \, ,
\end{align*}
which corresponds to Eq.\ (\ref{eq:NCASigma}) in Sec.\ \ref{sec:NCA}.

To perform actual calculations we work with a subset of Keldysh components \cite{Aoki:2014kx}, namely, 
the imaginary time Matsubara component $\Sp^M(\tau)$, 
the real-time greater component $\Sp^>(t, t')$,
the real-time lesser component $\Sp^<(t, t')$, 
and the right-mixing component $\Sp^\ttau(t, \tau')$.
These components can be derived from the general contour expression for $\Sp$ using the Langreth product rules \cite{Eckstein:2009qf, Stefanucci:2013oq}.
This is because the pseudo-particle self energy $\Sp$ is given by contour time products of the hybridization function $\bDelta$ and the pseudo-particle propagator $\Gp$, $\Sp \propto \bDelta \Gp$. Note that the two contributions $\Sp^{(1)}$ and $\Sp^{(2)}$ [Eqs.\ (\ref{eq:NCASigma1}) and (\ref{eq:NCASigma1})] must be treated differently as the order of the time arguments in $\bDelta$ differs. 

\subsection{Single particle Green's function}
\label{app:NCAGsp}

The NCA approximation for the single particle Green's function is given by the pseudo-particle bubble equipped with two vertices \cite{Eckstein:2010fk}. 
The Nambu generalization amounts to adding Nambu indices to the vertices and gives the non-connected Green's function as
\begin{multline}
  \tilde{\bG}_{\gamma \nu}(\ct, \ct') = (i) \times 
  \Tr \Bigg[
  \includegraphics[valign=c]
    {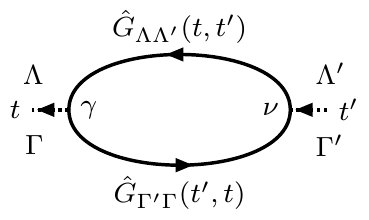}
  \Bigg]
  \\ = 
  i \Tr \big[ \hat{G}(\ct', \ct) \,\bba_\gamma (\ct) \, \hat{G}(\ct, \ct') \,\bbc_\nu(\ct') \big]
  \, ,
\end{multline}
where the hybridization line stubs denote the insertion of a Nambu creation or annihilation operator, and the trace corresponds to the summation over the $\Gamma'$ many-body state index.
To obtain the connected Green's function $\bG$ from $\tilde{\bG}$, the symmetry broken contribution must be removed, i.e.\ $\bG(\ct, \ct') = \tilde{\bG}(\ct, \ct') + i \Phi(\ct) \Phi^\dagger(\ct')$, which corresponds to Eq.\ (\ref{eq:NCAGsp}) and Fig.\ \ref{fig:NCA}b in Sec.\ \ref{sec:NCA}.

%

\bibliography{/Users/hugstr/Documents/Papers/DMFT_Biblography}

\begin{thebibliography}{81}%
\makeatletter
\providecommand \@ifxundefined [1]{%
 \@ifx{#1\undefined}
}%
\providecommand \@ifnum [1]{%
 \ifnum #1\expandafter \@firstoftwo
 \else \expandafter \@secondoftwo
 \fi
}%
\providecommand \@ifx [1]{%
 \ifx #1\expandafter \@firstoftwo
 \else \expandafter \@secondoftwo
 \fi
}%
\providecommand \natexlab [1]{#1}%
\providecommand \enquote  [1]{``#1''}%
\providecommand \bibnamefont  [1]{#1}%
\providecommand \bibfnamefont [1]{#1}%
\providecommand \citenamefont [1]{#1}%
\providecommand \href@noop [0]{\@secondoftwo}%
\providecommand \href [0]{\begingroup \@sanitize@url \@href}%
\providecommand \@href[1]{\@@startlink{#1}\@@href}%
\providecommand \@@href[1]{\endgroup#1\@@endlink}%
\providecommand \@sanitize@url [0]{\catcode `\\12\catcode `\$12\catcode
  `\&12\catcode `\#12\catcode `\^12\catcode `\_12\catcode `\%12\relax}%
\providecommand \@@startlink[1]{}%
\providecommand \@@endlink[0]{}%
\providecommand \url  [0]{\begingroup\@sanitize@url \@url }%
\providecommand \@url [1]{\endgroup\@href {#1}{\urlprefix }}%
\providecommand \urlprefix  [0]{URL }%
\providecommand \Eprint [0]{\href }%
\providecommand \doibase [0]{http://dx.doi.org/}%
\providecommand \selectlanguage [0]{\@gobble}%
\providecommand \bibinfo  [0]{\@secondoftwo}%
\providecommand \bibfield  [0]{\@secondoftwo}%
\providecommand \translation [1]{[#1]}%
\providecommand \BibitemOpen [0]{}%
\providecommand \bibitemStop [0]{}%
\providecommand \bibitemNoStop [0]{.\EOS\space}%
\providecommand \EOS [0]{\spacefactor3000\relax}%
\providecommand \BibitemShut  [1]{\csname bibitem#1\endcsname}%
\let\auto@bib@innerbib\@empty
\bibitem [{\citenamefont {Morsch}\ and\ \citenamefont
  {Oberthaler}(2006)}]{Morsch:2006vn}%
  \BibitemOpen
  \bibfield  {author} {\bibinfo {author} {\bibfnamefont {Oliver}\ \bibnamefont
  {Morsch}}\ and\ \bibinfo {author} {\bibfnamefont {Markus}\ \bibnamefont
  {Oberthaler}},\ }\bibfield  {title} {\enquote {\bibinfo {title} {Dynamics of
  bose-einstein condensates in optical lattices},}\ }\href {\doibase
  10.1103/RevModPhys.78.179} {\bibfield  {journal} {\bibinfo  {journal} {Rev.
  Mod. Phys.}\ }\textbf {\bibinfo {volume} {78}},\ \bibinfo {pages} {179--215}
  (\bibinfo {year} {2006})}\BibitemShut {NoStop}%
\bibitem [{\citenamefont {Bloch}\ \emph {et~al.}(2008)\citenamefont {Bloch},
  \citenamefont {Dalibard},\ and\ \citenamefont {Zwerger}}]{Bloch:2008uq}%
  \BibitemOpen
  \bibfield  {author} {\bibinfo {author} {\bibfnamefont {Immanuel}\
  \bibnamefont {Bloch}}, \bibinfo {author} {\bibfnamefont {Jean}\ \bibnamefont
  {Dalibard}}, \ and\ \bibinfo {author} {\bibfnamefont {Wilhelm}\ \bibnamefont
  {Zwerger}},\ }\bibfield  {title} {\enquote {\bibinfo {title} {Many-body
  physics with ultracold gases},}\ }\href {\doibase 10.1103/RevModPhys.80.885}
  {\bibfield  {journal} {\bibinfo  {journal} {Rev. Mod. Phys.}\ }\textbf
  {\bibinfo {volume} {80}},\ \bibinfo {pages} {885--964} (\bibinfo {year}
  {2008})}\BibitemShut {NoStop}%
\bibitem [{\citenamefont {Hubbard}(1963)}]{Hubbard:1963aa}%
  \BibitemOpen
  \bibfield  {author} {\bibinfo {author} {\bibfnamefont {J.}~\bibnamefont
  {Hubbard}},\ }\bibfield  {title} {\enquote {\bibinfo {title} {Electron
  correlations in narrow energy bands},}\ }\href@noop {} {\bibfield  {journal}
  {\bibinfo  {journal} {Proc. R. Soc. Lon. Ser.-A}\ }\textbf {\bibinfo {volume}
  {276}},\ \bibinfo {pages} {238--257} (\bibinfo {year} {1963})}\BibitemShut
  {NoStop}%
\bibitem [{\citenamefont {Fisher}\ \emph {et~al.}(1989)\citenamefont {Fisher},
  \citenamefont {Weichman}, \citenamefont {Grinstein},\ and\ \citenamefont
  {Fisher}}]{Fisher:1989kl}%
  \BibitemOpen
  \bibfield  {author} {\bibinfo {author} {\bibfnamefont {Matthew P.~A.}\
  \bibnamefont {Fisher}}, \bibinfo {author} {\bibfnamefont {Peter~B.}\
  \bibnamefont {Weichman}}, \bibinfo {author} {\bibfnamefont {G.}~\bibnamefont
  {Grinstein}}, \ and\ \bibinfo {author} {\bibfnamefont {Daniel~S.}\
  \bibnamefont {Fisher}},\ }\bibfield  {title} {\enquote {\bibinfo {title}
  {Boson localization and the superfluid-insulator transition},}\ }\href
  {\doibase 10.1103/PhysRevB.40.546} {\bibfield  {journal} {\bibinfo  {journal}
  {Phys. Rev. B}\ }\textbf {\bibinfo {volume} {40}},\ \bibinfo {pages}
  {546--570} (\bibinfo {year} {1989})}\BibitemShut {NoStop}%
\bibitem [{\citenamefont {Jaksch}\ \emph {et~al.}(1998)\citenamefont {Jaksch},
  \citenamefont {Bruder}, \citenamefont {Cirac}, \citenamefont {Gardiner},\
  and\ \citenamefont {Zoller}}]{Jaksch:1998vn}%
  \BibitemOpen
  \bibfield  {author} {\bibinfo {author} {\bibfnamefont {D.}~\bibnamefont
  {Jaksch}}, \bibinfo {author} {\bibfnamefont {C.}~\bibnamefont {Bruder}},
  \bibinfo {author} {\bibfnamefont {J.~I.}\ \bibnamefont {Cirac}}, \bibinfo
  {author} {\bibfnamefont {C.~W.}\ \bibnamefont {Gardiner}}, \ and\ \bibinfo
  {author} {\bibfnamefont {P.}~\bibnamefont {Zoller}},\ }\bibfield  {title}
  {\enquote {\bibinfo {title} {Cold bosonic atoms in optical lattices},}\
  }\href {\doibase 10.1103/PhysRevLett.81.3108} {\bibfield  {journal} {\bibinfo
   {journal} {Phys. Rev. Lett.}\ }\textbf {\bibinfo {volume} {81}},\ \bibinfo
  {pages} {3108--3111} (\bibinfo {year} {1998})}\BibitemShut {NoStop}%
\bibitem [{\citenamefont {Trotzky}\ \emph {et~al.}(2010)\citenamefont
  {Trotzky}, \citenamefont {Pollet}, \citenamefont {Gerbier}, \citenamefont
  {Schnorrberger}, \citenamefont {Bloch}, \citenamefont {Prokof\'ev},
  \citenamefont {Svistunov},\ and\ \citenamefont {Troyer}}]{Trotzky:2010fk}%
  \BibitemOpen
  \bibfield  {author} {\bibinfo {author} {\bibfnamefont {S.}~\bibnamefont
  {Trotzky}}, \bibinfo {author} {\bibfnamefont {L.}~\bibnamefont {Pollet}},
  \bibinfo {author} {\bibfnamefont {F.}~\bibnamefont {Gerbier}}, \bibinfo
  {author} {\bibfnamefont {U.}~\bibnamefont {Schnorrberger}}, \bibinfo {author}
  {\bibfnamefont {I.}~\bibnamefont {Bloch}}, \bibinfo {author} {\bibfnamefont
  {N.~V.}\ \bibnamefont {Prokof\'ev}}, \bibinfo {author} {\bibfnamefont
  {B.}~\bibnamefont {Svistunov}}, \ and\ \bibinfo {author} {\bibfnamefont
  {M.}~\bibnamefont {Troyer}},\ }\bibfield  {title} {\enquote {\bibinfo {title}
  {Suppression of the critical temperature for superfluidity near the mott
  transition},}\ }\href@noop {} {\bibfield  {journal} {\bibinfo  {journal}
  {Nat. Phys.}\ }\textbf {\bibinfo {volume} {6}},\ \bibinfo {pages} {998--1004}
  (\bibinfo {year} {2010})}\BibitemShut {NoStop}%
\bibitem [{\citenamefont {Kennett}(2013)}]{Kennett:2013fk}%
  \BibitemOpen
  \bibfield  {author} {\bibinfo {author} {\bibfnamefont {M.~P.}\ \bibnamefont
  {Kennett}},\ }\bibfield  {title} {\enquote {\bibinfo {title}
  {Out-of-equilibrium dynamics of the bose-hubbard model},}\ }\href@noop {}
  {\bibfield  {journal} {\bibinfo  {journal} {ISRN Cond. Mat. Phys.}\ }\textbf
  {\bibinfo {volume} {2013}},\ \bibinfo {pages} {39} (\bibinfo {year}
  {2013})}\BibitemShut {NoStop}%
\bibitem [{\citenamefont {Greiner}\ \emph {et~al.}(2002)\citenamefont
  {Greiner}, \citenamefont {Mandel}, \citenamefont {Hansch},\ and\
  \citenamefont {Bloch}}]{Greiner:2002fk}%
  \BibitemOpen
  \bibfield  {author} {\bibinfo {author} {\bibfnamefont {Markus}\ \bibnamefont
  {Greiner}}, \bibinfo {author} {\bibfnamefont {Olaf}\ \bibnamefont {Mandel}},
  \bibinfo {author} {\bibfnamefont {Theodor~W.}\ \bibnamefont {Hansch}}, \ and\
  \bibinfo {author} {\bibfnamefont {Immanuel}\ \bibnamefont {Bloch}},\
  }\bibfield  {title} {\enquote {\bibinfo {title} {Collapse and revival of the
  matter wave field of a bose-einstein condensate},}\ }\href@noop {} {\bibfield
   {journal} {\bibinfo  {journal} {Nature}\ }\textbf {\bibinfo {volume}
  {419}},\ \bibinfo {pages} {51--54} (\bibinfo {year} {2002})}\BibitemShut
  {NoStop}%
\bibitem [{\citenamefont {Sebby-Strabley}\ \emph {et~al.}(2007)\citenamefont
  {Sebby-Strabley}, \citenamefont {Brown}, \citenamefont {Anderlini},
  \citenamefont {Lee}, \citenamefont {Phillips}, \citenamefont {Porto},\ and\
  \citenamefont {Johnson}}]{Sebby-Strabley:2007vn}%
  \BibitemOpen
  \bibfield  {author} {\bibinfo {author} {\bibfnamefont {J.}~\bibnamefont
  {Sebby-Strabley}}, \bibinfo {author} {\bibfnamefont {B.~L.}\ \bibnamefont
  {Brown}}, \bibinfo {author} {\bibfnamefont {M.}~\bibnamefont {Anderlini}},
  \bibinfo {author} {\bibfnamefont {P.~J.}\ \bibnamefont {Lee}}, \bibinfo
  {author} {\bibfnamefont {W.~D.}\ \bibnamefont {Phillips}}, \bibinfo {author}
  {\bibfnamefont {J.~V.}\ \bibnamefont {Porto}}, \ and\ \bibinfo {author}
  {\bibfnamefont {P.~R.}\ \bibnamefont {Johnson}},\ }\bibfield  {title}
  {\enquote {\bibinfo {title} {Preparing and probing atomic number states with
  an atom interferometer},}\ }\href {\doibase 10.1103/PhysRevLett.98.200405}
  {\bibfield  {journal} {\bibinfo  {journal} {Phys. Rev. Lett.}\ }\textbf
  {\bibinfo {volume} {98}},\ \bibinfo {pages} {200405} (\bibinfo {year}
  {2007})}\BibitemShut {NoStop}%
\bibitem [{\citenamefont {Will}\ \emph {et~al.}(2010)\citenamefont {Will},
  \citenamefont {Best}, \citenamefont {Schneider}, \citenamefont
  {Hackermuller}, \citenamefont {Luhmann},\ and\ \citenamefont
  {Bloch}}]{Will:2010uq}%
  \BibitemOpen
  \bibfield  {author} {\bibinfo {author} {\bibfnamefont {Sebastian}\
  \bibnamefont {Will}}, \bibinfo {author} {\bibfnamefont {Thorsten}\
  \bibnamefont {Best}}, \bibinfo {author} {\bibfnamefont {Ulrich}\ \bibnamefont
  {Schneider}}, \bibinfo {author} {\bibfnamefont {Lucia}\ \bibnamefont
  {Hackermuller}}, \bibinfo {author} {\bibfnamefont {Dirk-Soren}\ \bibnamefont
  {Luhmann}}, \ and\ \bibinfo {author} {\bibfnamefont {Immanuel}\ \bibnamefont
  {Bloch}},\ }\bibfield  {title} {\enquote {\bibinfo {title} {Time-resolved
  observation of coherent multi-body interactions in quantum phase revivals},}\
  }\href@noop {} {\bibfield  {journal} {\bibinfo  {journal} {Nature}\ }\textbf
  {\bibinfo {volume} {465}},\ \bibinfo {pages} {197--201} (\bibinfo {year}
  {2010})}\BibitemShut {NoStop}%
\bibitem [{\citenamefont {Bakr}\ \emph {et~al.}(2010)\citenamefont {Bakr},
  \citenamefont {Peng}, \citenamefont {Tai}, \citenamefont {Ma}, \citenamefont
  {Simon}, \citenamefont {Gillen}, \citenamefont {F{\"o}lling}, \citenamefont
  {Pollet},\ and\ \citenamefont {Greiner}}]{Bakr:2010dq}%
  \BibitemOpen
  \bibfield  {author} {\bibinfo {author} {\bibfnamefont {W.~S.}\ \bibnamefont
  {Bakr}}, \bibinfo {author} {\bibfnamefont {A.}~\bibnamefont {Peng}}, \bibinfo
  {author} {\bibfnamefont {M.~E.}\ \bibnamefont {Tai}}, \bibinfo {author}
  {\bibfnamefont {R.}~\bibnamefont {Ma}}, \bibinfo {author} {\bibfnamefont
  {J.}~\bibnamefont {Simon}}, \bibinfo {author} {\bibfnamefont {J.~I.}\
  \bibnamefont {Gillen}}, \bibinfo {author} {\bibfnamefont {S.}~\bibnamefont
  {F{\"o}lling}}, \bibinfo {author} {\bibfnamefont {L.}~\bibnamefont {Pollet}},
  \ and\ \bibinfo {author} {\bibfnamefont {M.}~\bibnamefont {Greiner}},\
  }\bibfield  {title} {\enquote {\bibinfo {title} {Probing the
  superfluid--to--mott insulator transition at the single-atom level},}\
  }\href@noop {} {\bibfield  {journal} {\bibinfo  {journal} {Science}\ }\textbf
  {\bibinfo {volume} {329}},\ \bibinfo {pages} {547--550} (\bibinfo {year}
  {2010})}\BibitemShut {NoStop}%
\bibitem [{\citenamefont {Bissbort}\ \emph {et~al.}(2011)\citenamefont
  {Bissbort}, \citenamefont {G\"otze}, \citenamefont {Li}, \citenamefont
  {Heinze}, \citenamefont {Krauser}, \citenamefont {Weinberg}, \citenamefont
  {Becker}, \citenamefont {Sengstock},\ and\ \citenamefont
  {Hofstetter}}]{Bissbort:2011bs}%
  \BibitemOpen
  \bibfield  {author} {\bibinfo {author} {\bibfnamefont {Ulf}\ \bibnamefont
  {Bissbort}}, \bibinfo {author} {\bibfnamefont {S\"oren}\ \bibnamefont
  {G\"otze}}, \bibinfo {author} {\bibfnamefont {Yongqiang}\ \bibnamefont {Li}},
  \bibinfo {author} {\bibfnamefont {Jannes}\ \bibnamefont {Heinze}}, \bibinfo
  {author} {\bibfnamefont {Jasper~S.}\ \bibnamefont {Krauser}}, \bibinfo
  {author} {\bibfnamefont {Malte}\ \bibnamefont {Weinberg}}, \bibinfo {author}
  {\bibfnamefont {Christoph}\ \bibnamefont {Becker}}, \bibinfo {author}
  {\bibfnamefont {Klaus}\ \bibnamefont {Sengstock}}, \ and\ \bibinfo {author}
  {\bibfnamefont {Walter}\ \bibnamefont {Hofstetter}},\ }\bibfield  {title}
  {\enquote {\bibinfo {title} {Detecting the amplitude mode of strongly
  interacting lattice bosons by bragg scattering},}\ }\href {\doibase
  10.1103/PhysRevLett.106.205303} {\bibfield  {journal} {\bibinfo  {journal}
  {Phys. Rev. Lett.}\ }\textbf {\bibinfo {volume} {106}},\ \bibinfo {pages}
  {205303} (\bibinfo {year} {2011})}\BibitemShut {NoStop}%
\bibitem [{\citenamefont {Endres}\ \emph {et~al.}(2012)\citenamefont {Endres},
  \citenamefont {Fukuhara}, \citenamefont {Pekker}, \citenamefont {Cheneau},
  \citenamefont {Schau\ss}, \citenamefont {Gross}, \citenamefont {Demler},
  \citenamefont {Kuhr},\ and\ \citenamefont {Bloch}}]{Endres:2012bs}%
  \BibitemOpen
  \bibfield  {author} {\bibinfo {author} {\bibfnamefont {Manuel}\ \bibnamefont
  {Endres}}, \bibinfo {author} {\bibfnamefont {Takeshi}\ \bibnamefont
  {Fukuhara}}, \bibinfo {author} {\bibfnamefont {David}\ \bibnamefont
  {Pekker}}, \bibinfo {author} {\bibfnamefont {Marc}\ \bibnamefont {Cheneau}},
  \bibinfo {author} {\bibfnamefont {Peter}\ \bibnamefont {Schau\ss}}, \bibinfo
  {author} {\bibfnamefont {Christian}\ \bibnamefont {Gross}}, \bibinfo {author}
  {\bibfnamefont {Eugene}\ \bibnamefont {Demler}}, \bibinfo {author}
  {\bibfnamefont {Stefan}\ \bibnamefont {Kuhr}}, \ and\ \bibinfo {author}
  {\bibfnamefont {Immanuel}\ \bibnamefont {Bloch}},\ }\bibfield  {title}
  {\enquote {\bibinfo {title} {The /`higgs/' amplitude mode at the
  two-dimensional superfluid/mott insulator transition},}\ }\href@noop {}
  {\bibfield  {journal} {\bibinfo  {journal} {Nature}\ }\textbf {\bibinfo
  {volume} {487}},\ \bibinfo {pages} {454--458} (\bibinfo {year}
  {2012})}\BibitemShut {NoStop}%
\bibitem [{\citenamefont {Cheneau}\ \emph {et~al.}(2012)\citenamefont
  {Cheneau}, \citenamefont {Barmettler}, \citenamefont {Poletti}, \citenamefont
  {Endres}, \citenamefont {Schausz}, \citenamefont {Fukuhara}, \citenamefont
  {Gross}, \citenamefont {Bloch}, \citenamefont {Kollath},\ and\ \citenamefont
  {Kuhr}}]{Cheneau:2012ys}%
  \BibitemOpen
  \bibfield  {author} {\bibinfo {author} {\bibfnamefont {Marc}\ \bibnamefont
  {Cheneau}}, \bibinfo {author} {\bibfnamefont {Peter}\ \bibnamefont
  {Barmettler}}, \bibinfo {author} {\bibfnamefont {Dario}\ \bibnamefont
  {Poletti}}, \bibinfo {author} {\bibfnamefont {Manuel}\ \bibnamefont
  {Endres}}, \bibinfo {author} {\bibfnamefont {Peter}\ \bibnamefont {Schausz}},
  \bibinfo {author} {\bibfnamefont {Takeshi}\ \bibnamefont {Fukuhara}},
  \bibinfo {author} {\bibfnamefont {Christian}\ \bibnamefont {Gross}}, \bibinfo
  {author} {\bibfnamefont {Immanuel}\ \bibnamefont {Bloch}}, \bibinfo {author}
  {\bibfnamefont {Corinna}\ \bibnamefont {Kollath}}, \ and\ \bibinfo {author}
  {\bibfnamefont {Stefan}\ \bibnamefont {Kuhr}},\ }\bibfield  {title} {\enquote
  {\bibinfo {title} {Light-cone-like spreading of correlations in a quantum
  many-body system},}\ }\href@noop {} {\bibfield  {journal} {\bibinfo
  {journal} {Nature}\ }\textbf {\bibinfo {volume} {481}},\ \bibinfo {pages}
  {484--487} (\bibinfo {year} {2012})}\BibitemShut {NoStop}%
\bibitem [{\citenamefont {Trotzky}\ \emph {et~al.}(2012)\citenamefont
  {Trotzky}, \citenamefont {Chen}, \citenamefont {Flesch}, \citenamefont
  {McCulloch}, \citenamefont {Schollwock}, \citenamefont {Eisert},\ and\
  \citenamefont {Bloch}}]{Trotzky:2012kx}%
  \BibitemOpen
  \bibfield  {author} {\bibinfo {author} {\bibfnamefont {S.}~\bibnamefont
  {Trotzky}}, \bibinfo {author} {\bibfnamefont {Y-A.}\ \bibnamefont {Chen}},
  \bibinfo {author} {\bibfnamefont {A.}~\bibnamefont {Flesch}}, \bibinfo
  {author} {\bibfnamefont {I.~P.}\ \bibnamefont {McCulloch}}, \bibinfo {author}
  {\bibfnamefont {U.}~\bibnamefont {Schollwock}}, \bibinfo {author}
  {\bibfnamefont {J.}~\bibnamefont {Eisert}}, \ and\ \bibinfo {author}
  {\bibfnamefont {I.}~\bibnamefont {Bloch}},\ }\bibfield  {title} {\enquote
  {\bibinfo {title} {Probing the relaxation towards equilibrium in an isolated
  strongly correlated one-dimensional bose gas},}\ }\href@noop {} {\bibfield
  {journal} {\bibinfo  {journal} {Nat. Phys.}\ }\textbf {\bibinfo {volume}
  {8}},\ \bibinfo {pages} {325--330} (\bibinfo {year} {2012})}\BibitemShut
  {NoStop}%
\bibitem [{\citenamefont {{Braun}}\ \emph {et~al.}(2014)\citenamefont
  {{Braun}}, \citenamefont {{Friesdorf}}, \citenamefont {{Hodgman}},
  \citenamefont {{Schreiber}}, \citenamefont {{Ronzheimer}}, \citenamefont
  {{Riera}}, \citenamefont {{del Rey}}, \citenamefont {{Bloch}}, \citenamefont
  {{Eisert}},\ and\ \citenamefont {{Schneider}}}]{Braun:2014kx}%
  \BibitemOpen
  \bibfield  {author} {\bibinfo {author} {\bibfnamefont {S.}~\bibnamefont
  {{Braun}}}, \bibinfo {author} {\bibfnamefont {M.}~\bibnamefont
  {{Friesdorf}}}, \bibinfo {author} {\bibfnamefont {S.~S.}\ \bibnamefont
  {{Hodgman}}}, \bibinfo {author} {\bibfnamefont {M.}~\bibnamefont
  {{Schreiber}}}, \bibinfo {author} {\bibfnamefont {J.~P.}\ \bibnamefont
  {{Ronzheimer}}}, \bibinfo {author} {\bibfnamefont {A.}~\bibnamefont
  {{Riera}}}, \bibinfo {author} {\bibfnamefont {M.}~\bibnamefont {{del Rey}}},
  \bibinfo {author} {\bibfnamefont {I.}~\bibnamefont {{Bloch}}}, \bibinfo
  {author} {\bibfnamefont {J.}~\bibnamefont {{Eisert}}}, \ and\ \bibinfo
  {author} {\bibfnamefont {U.}~\bibnamefont {{Schneider}}},\ }\bibfield
  {title} {\enquote {\bibinfo {title} {{Emergence of coherence and the dynamics
  of quantum phase transitions}},}\ }\href@noop {} {\bibfield  {journal}
  {\bibinfo  {journal} {ArXiv e-prints}\ } (\bibinfo {year} {2014})},\ \Eprint
  {http://arxiv.org/abs/1403.7199} {arXiv:1403.7199 [cond-mat.quant-gas]}
  \BibitemShut {NoStop}%
\bibitem [{\citenamefont {Capogrosso-Sansone}\ \emph
  {et~al.}(2007)\citenamefont {Capogrosso-Sansone}, \citenamefont {Prokof'ev},\
  and\ \citenamefont {Svistunov}}]{Capogrosso-Sansone:2007lh}%
  \BibitemOpen
  \bibfield  {author} {\bibinfo {author} {\bibfnamefont {B.}~\bibnamefont
  {Capogrosso-Sansone}}, \bibinfo {author} {\bibfnamefont {N.~V.}\ \bibnamefont
  {Prokof'ev}}, \ and\ \bibinfo {author} {\bibfnamefont {B.~V.}\ \bibnamefont
  {Svistunov}},\ }\bibfield  {title} {\enquote {\bibinfo {title} {Phase diagram
  and thermodynamics of the three-dimensional bose-hubbard model},}\ }\href
  {\doibase 10.1103/PhysRevB.75.134302} {\bibfield  {journal} {\bibinfo
  {journal} {Phys. Rev. B}\ }\textbf {\bibinfo {volume} {75}},\ \bibinfo
  {pages} {134302} (\bibinfo {year} {2007})}\BibitemShut {NoStop}%
\bibitem [{\citenamefont {Pollet}(2012)}]{Pollet:2012ly}%
  \BibitemOpen
  \bibfield  {author} {\bibinfo {author} {\bibfnamefont {Lode}\ \bibnamefont
  {Pollet}},\ }\bibfield  {title} {\enquote {\bibinfo {title} {Recent
  developments in quantum monte carlo simulations with applications for cold
  gases},}\ }\href@noop {} {\bibfield  {journal} {\bibinfo  {journal} {Rep.
  Prog. Phys.}\ }\textbf {\bibinfo {volume} {75}},\ \bibinfo {pages} {094501}
  (\bibinfo {year} {2012})}\BibitemShut {NoStop}%
\bibitem [{\citenamefont {Schollw\"ock}(2005)}]{Schollwock:2005ly}%
  \BibitemOpen
  \bibfield  {author} {\bibinfo {author} {\bibfnamefont {U.}~\bibnamefont
  {Schollw\"ock}},\ }\bibfield  {title} {\enquote {\bibinfo {title} {The
  density-matrix renormalization group},}\ }\href {\doibase
  10.1103/RevModPhys.77.259} {\bibfield  {journal} {\bibinfo  {journal} {Rev.
  Mod. Phys.}\ }\textbf {\bibinfo {volume} {77}},\ \bibinfo {pages} {259--315}
  (\bibinfo {year} {2005})}\BibitemShut {NoStop}%
\bibitem [{\citenamefont {Kollath}\ \emph {et~al.}(2007)\citenamefont
  {Kollath}, \citenamefont {L\"auchli},\ and\ \citenamefont
  {Altman}}]{Kollath:2007ys}%
  \BibitemOpen
  \bibfield  {author} {\bibinfo {author} {\bibfnamefont {Corinna}\ \bibnamefont
  {Kollath}}, \bibinfo {author} {\bibfnamefont {Andreas~M.}\ \bibnamefont
  {L\"auchli}}, \ and\ \bibinfo {author} {\bibfnamefont {Ehud}\ \bibnamefont
  {Altman}},\ }\bibfield  {title} {\enquote {\bibinfo {title} {Quench dynamics
  and nonequilibrium phase diagram of the bose-hubbard model},}\ }\href
  {\doibase 10.1103/PhysRevLett.98.180601} {\bibfield  {journal} {\bibinfo
  {journal} {Phys. Rev. Lett.}\ }\textbf {\bibinfo {volume} {98}},\ \bibinfo
  {pages} {180601} (\bibinfo {year} {2007})}\BibitemShut {NoStop}%
\bibitem [{\citenamefont {Roux}(2009)}]{Roux:2009ys}%
  \BibitemOpen
  \bibfield  {author} {\bibinfo {author} {\bibfnamefont {Guillaume}\
  \bibnamefont {Roux}},\ }\bibfield  {title} {\enquote {\bibinfo {title}
  {Quenches in quantum many-body systems: One-dimensional bose-hubbard model
  reexamined},}\ }\href {\doibase 10.1103/PhysRevA.79.021608} {\bibfield
  {journal} {\bibinfo  {journal} {Phys. Rev. A}\ }\textbf {\bibinfo {volume}
  {79}},\ \bibinfo {pages} {021608} (\bibinfo {year} {2009})}\BibitemShut
  {NoStop}%
\bibitem [{\citenamefont {Sorg}\ \emph {et~al.}(2014)\citenamefont {Sorg},
  \citenamefont {Vidmar}, \citenamefont {Pollet},\ and\ \citenamefont
  {Heidrich-Meisner}}]{Sorg:2014vn}%
  \BibitemOpen
  \bibfield  {author} {\bibinfo {author} {\bibfnamefont {S.}~\bibnamefont
  {Sorg}}, \bibinfo {author} {\bibfnamefont {L.}~\bibnamefont {Vidmar}},
  \bibinfo {author} {\bibfnamefont {L.}~\bibnamefont {Pollet}}, \ and\ \bibinfo
  {author} {\bibfnamefont {F.}~\bibnamefont {Heidrich-Meisner}},\ }\bibfield
  {title} {\enquote {\bibinfo {title} {Relaxation and thermalization in the
  one-dimensional bose-hubbard model: A case study for the interaction quantum
  quench from the atomic limit},}\ }\href {\doibase 10.1103/PhysRevA.90.033606}
  {\bibfield  {journal} {\bibinfo  {journal} {Phys. Rev. A}\ }\textbf {\bibinfo
  {volume} {90}},\ \bibinfo {pages} {033606} (\bibinfo {year}
  {2014})}\BibitemShut {NoStop}%
\bibitem [{\citenamefont {Rigol}(2010)}]{Rigol:2010zr}%
  \BibitemOpen
  \bibfield  {author} {\bibinfo {author} {\bibfnamefont {Marcos}\ \bibnamefont
  {Rigol}},\ }\bibfield  {title} {\enquote {\bibinfo {title} {Comment on
  ``quenches in quantum many-body systems: One-dimensional bose-hubbard model
  reexamined''},}\ }\href {\doibase 10.1103/PhysRevA.82.037601} {\bibfield
  {journal} {\bibinfo  {journal} {Phys. Rev. A}\ }\textbf {\bibinfo {volume}
  {82}},\ \bibinfo {pages} {037601} (\bibinfo {year} {2010})}\BibitemShut
  {NoStop}%
\bibitem [{\citenamefont {Roux}(2010)}]{Roux:2010ly}%
  \BibitemOpen
  \bibfield  {author} {\bibinfo {author} {\bibfnamefont {Guillaume}\
  \bibnamefont {Roux}},\ }\bibfield  {title} {\enquote {\bibinfo {title} {Reply
  to ``comment on `quenches in quantum many-body systems: One-dimensional
  bose-hubbard model reexamined' ''},}\ }\href {\doibase
  10.1103/PhysRevA.82.037602} {\bibfield  {journal} {\bibinfo  {journal} {Phys.
  Rev. A}\ }\textbf {\bibinfo {volume} {82}},\ \bibinfo {pages} {037602}
  (\bibinfo {year} {2010})}\BibitemShut {NoStop}%
\bibitem [{\citenamefont {Carleo}\ \emph {et~al.}(2012)\citenamefont {Carleo},
  \citenamefont {Becca}, \citenamefont {Schir{\'o}},\ and\ \citenamefont
  {Fabrizio}}]{Carleo:2012zr}%
  \BibitemOpen
  \bibfield  {author} {\bibinfo {author} {\bibfnamefont {Giuseppe}\
  \bibnamefont {Carleo}}, \bibinfo {author} {\bibfnamefont {Federico}\
  \bibnamefont {Becca}}, \bibinfo {author} {\bibfnamefont {Marco}\ \bibnamefont
  {Schir{\'o}}}, \ and\ \bibinfo {author} {\bibfnamefont {Michele}\
  \bibnamefont {Fabrizio}},\ }\bibfield  {title} {\enquote {\bibinfo {title}
  {Localization and glassy dynamics of many-body quantum systems},}\
  }\href@noop {} {\bibfield  {journal} {\bibinfo  {journal} {Sci. Rep.}\
  }\textbf {\bibinfo {volume} {2}} (\bibinfo {year} {2012})}\BibitemShut
  {NoStop}%
\bibitem [{\citenamefont {Carleo}\ \emph {et~al.}(2014)\citenamefont {Carleo},
  \citenamefont {Becca}, \citenamefont {Sanchez-Palencia}, \citenamefont
  {Sorella},\ and\ \citenamefont {Fabrizio}}]{Carleo:2014ly}%
  \BibitemOpen
  \bibfield  {author} {\bibinfo {author} {\bibfnamefont {Giuseppe}\
  \bibnamefont {Carleo}}, \bibinfo {author} {\bibfnamefont {Federico}\
  \bibnamefont {Becca}}, \bibinfo {author} {\bibfnamefont {Laurent}\
  \bibnamefont {Sanchez-Palencia}}, \bibinfo {author} {\bibfnamefont {Sandro}\
  \bibnamefont {Sorella}}, \ and\ \bibinfo {author} {\bibfnamefont {Michele}\
  \bibnamefont {Fabrizio}},\ }\bibfield  {title} {\enquote {\bibinfo {title}
  {Light-cone effect and supersonic correlations in one- and two-dimensional
  bosonic superfluids},}\ }\href {\doibase 10.1103/PhysRevA.89.031602}
  {\bibfield  {journal} {\bibinfo  {journal} {Phys. Rev. A}\ }\textbf {\bibinfo
  {volume} {89}},\ \bibinfo {pages} {031602} (\bibinfo {year}
  {2014})}\BibitemShut {NoStop}%
\bibitem [{\citenamefont {Biroli}\ \emph {et~al.}(2010)\citenamefont {Biroli},
  \citenamefont {Kollath},\ and\ \citenamefont {L\"auchli}}]{Biroli:2010qf}%
  \BibitemOpen
  \bibfield  {author} {\bibinfo {author} {\bibfnamefont {Giulio}\ \bibnamefont
  {Biroli}}, \bibinfo {author} {\bibfnamefont {Corinna}\ \bibnamefont
  {Kollath}}, \ and\ \bibinfo {author} {\bibfnamefont {Andreas~M.}\
  \bibnamefont {L\"auchli}},\ }\bibfield  {title} {\enquote {\bibinfo {title}
  {Effect of rare fluctuations on the thermalization of isolated quantum
  systems},}\ }\href {\doibase 10.1103/PhysRevLett.105.250401} {\bibfield
  {journal} {\bibinfo  {journal} {Phys. Rev. Lett.}\ }\textbf {\bibinfo
  {volume} {105}},\ \bibinfo {pages} {250401} (\bibinfo {year}
  {2010})}\BibitemShut {NoStop}%
\bibitem [{\citenamefont {Trefzger}\ and\ \citenamefont
  {Sengupta}(2011)}]{Trefzger:2011uq}%
  \BibitemOpen
  \bibfield  {author} {\bibinfo {author} {\bibfnamefont {C.}~\bibnamefont
  {Trefzger}}\ and\ \bibinfo {author} {\bibfnamefont {K.}~\bibnamefont
  {Sengupta}},\ }\bibfield  {title} {\enquote {\bibinfo {title} {Nonequilibrium
  dynamics of the bose-hubbard model: A projection-operator approach},}\ }\href
  {\doibase 10.1103/PhysRevLett.106.095702} {\bibfield  {journal} {\bibinfo
  {journal} {Phys. Rev. Lett.}\ }\textbf {\bibinfo {volume} {106}},\ \bibinfo
  {pages} {095702} (\bibinfo {year} {2011})}\BibitemShut {NoStop}%
\bibitem [{\citenamefont {Kennett}\ and\ \citenamefont
  {Dalidovich}(2011)}]{Kennett:2011fk}%
  \BibitemOpen
  \bibfield  {author} {\bibinfo {author} {\bibfnamefont {Malcolm~P.}\
  \bibnamefont {Kennett}}\ and\ \bibinfo {author} {\bibfnamefont {Denis}\
  \bibnamefont {Dalidovich}},\ }\bibfield  {title} {\enquote {\bibinfo {title}
  {Schwinger-keldysh approach to out-of-equilibrium dynamics of the
  bose-hubbard model with time-varying hopping},}\ }\href {\doibase
  10.1103/PhysRevA.84.033620} {\bibfield  {journal} {\bibinfo  {journal} {Phys.
  Rev. A}\ }\textbf {\bibinfo {volume} {84}},\ \bibinfo {pages} {033620}
  (\bibinfo {year} {2011})}\BibitemShut {NoStop}%
\bibitem [{\citenamefont {Dutta}\ \emph {et~al.}(2012)\citenamefont {Dutta},
  \citenamefont {Trefzger},\ and\ \citenamefont {Sengupta}}]{Dutta:2012kx}%
  \BibitemOpen
  \bibfield  {author} {\bibinfo {author} {\bibfnamefont {Anirban}\ \bibnamefont
  {Dutta}}, \bibinfo {author} {\bibfnamefont {C.}~\bibnamefont {Trefzger}}, \
  and\ \bibinfo {author} {\bibfnamefont {K.}~\bibnamefont {Sengupta}},\
  }\bibfield  {title} {\enquote {\bibinfo {title} {Projection operator approach
  to the bose-hubbard model},}\ }\href {\doibase 10.1103/PhysRevB.86.085140}
  {\bibfield  {journal} {\bibinfo  {journal} {Phys. Rev. B}\ }\textbf {\bibinfo
  {volume} {86}},\ \bibinfo {pages} {085140} (\bibinfo {year}
  {2012})}\BibitemShut {NoStop}%
\bibitem [{\citenamefont {{Dutta}}\ \emph {et~al.}(2014)\citenamefont
  {{Dutta}}, \citenamefont {{Sensarma}},\ and\ \citenamefont
  {{Sengupta}}}]{Dutta:2014uq}%
  \BibitemOpen
  \bibfield  {author} {\bibinfo {author} {\bibfnamefont {A.}~\bibnamefont
  {{Dutta}}}, \bibinfo {author} {\bibfnamefont {R.}~\bibnamefont {{Sensarma}}},
  \ and\ \bibinfo {author} {\bibfnamefont {K.}~\bibnamefont {{Sengupta}}},\
  }\bibfield  {title} {\enquote {\bibinfo {title} {{Role of trap-induced scales
  in non-equilibrium dynamics of strongly interacting trapped bosons}},}\
  }\href@noop {} {\bibfield  {journal} {\bibinfo  {journal} {ArXiv e-prints}\ }
  (\bibinfo {year} {2014})},\ \Eprint {http://arxiv.org/abs/1406.0849}
  {arXiv:1406.0849 [cond-mat.str-el]} \BibitemShut {NoStop}%
\bibitem [{\citenamefont {Huber}\ \emph {et~al.}(2007)\citenamefont {Huber},
  \citenamefont {Altman}, \citenamefont {B\"uchler},\ and\ \citenamefont
  {Blatter}}]{Huber:2007ys}%
  \BibitemOpen
  \bibfield  {author} {\bibinfo {author} {\bibfnamefont {S.~D.}\ \bibnamefont
  {Huber}}, \bibinfo {author} {\bibfnamefont {E.}~\bibnamefont {Altman}},
  \bibinfo {author} {\bibfnamefont {H.~P.}\ \bibnamefont {B\"uchler}}, \ and\
  \bibinfo {author} {\bibfnamefont {G.}~\bibnamefont {Blatter}},\ }\bibfield
  {title} {\enquote {\bibinfo {title} {Dynamical properties of ultracold bosons
  in an optical lattice},}\ }\href {\doibase 10.1103/PhysRevB.75.085106}
  {\bibfield  {journal} {\bibinfo  {journal} {Phys. Rev. B}\ }\textbf {\bibinfo
  {volume} {75}},\ \bibinfo {pages} {085106} (\bibinfo {year}
  {2007})}\BibitemShut {NoStop}%
\bibitem [{\citenamefont {Wolf}\ \emph {et~al.}(2010)\citenamefont {Wolf},
  \citenamefont {Hen},\ and\ \citenamefont {Rigol}}]{Wolf:2010bh}%
  \BibitemOpen
  \bibfield  {author} {\bibinfo {author} {\bibfnamefont {F.~Alexander}\
  \bibnamefont {Wolf}}, \bibinfo {author} {\bibfnamefont {Itay}\ \bibnamefont
  {Hen}}, \ and\ \bibinfo {author} {\bibfnamefont {Marcos}\ \bibnamefont
  {Rigol}},\ }\bibfield  {title} {\enquote {\bibinfo {title} {Collapse and
  revival oscillations as a probe for the tunneling amplitude in an ultracold
  bose gas},}\ }\href {\doibase 10.1103/PhysRevA.82.043601} {\bibfield
  {journal} {\bibinfo  {journal} {Phys. Rev. A}\ }\textbf {\bibinfo {volume}
  {82}},\ \bibinfo {pages} {043601} (\bibinfo {year} {2010})}\BibitemShut
  {NoStop}%
\bibitem [{\citenamefont {Sciolla}\ and\ \citenamefont
  {Biroli}(2010)}]{Sciolla:2010uq}%
  \BibitemOpen
  \bibfield  {author} {\bibinfo {author} {\bibfnamefont {Bruno}\ \bibnamefont
  {Sciolla}}\ and\ \bibinfo {author} {\bibfnamefont {Giulio}\ \bibnamefont
  {Biroli}},\ }\bibfield  {title} {\enquote {\bibinfo {title} {Quantum quenches
  and off-equilibrium dynamical transition in the infinite-dimensional
  bose-hubbard model},}\ }\href {\doibase 10.1103/PhysRevLett.105.220401}
  {\bibfield  {journal} {\bibinfo  {journal} {Phys. Rev. Lett.}\ }\textbf
  {\bibinfo {volume} {105}},\ \bibinfo {pages} {220401} (\bibinfo {year}
  {2010})}\BibitemShut {NoStop}%
\bibitem [{\citenamefont {Sciolla}\ and\ \citenamefont
  {Biroli}(2011)}]{Sciolla:2011kx}%
  \BibitemOpen
  \bibfield  {author} {\bibinfo {author} {\bibfnamefont {Bruno}\ \bibnamefont
  {Sciolla}}\ and\ \bibinfo {author} {\bibfnamefont {Giulio}\ \bibnamefont
  {Biroli}},\ }\bibfield  {title} {\enquote {\bibinfo {title} {Dynamical
  transitions and quantum quenches in mean-field models},}\ }\href@noop {}
  {\bibfield  {journal} {\bibinfo  {journal} {J. Stat. Mech: Theory Exp.}\
  }\textbf {\bibinfo {volume} {2011}},\ \bibinfo {pages} {P11003} (\bibinfo
  {year} {2011})}\BibitemShut {NoStop}%
\bibitem [{\citenamefont {Snoek}(2011)}]{Snoek:2011hc}%
  \BibitemOpen
  \bibfield  {author} {\bibinfo {author} {\bibfnamefont {M.}~\bibnamefont
  {Snoek}},\ }\bibfield  {title} {\enquote {\bibinfo {title} {Rigorous
  mean-field dynamics of lattice bosons: Quenches from the mott insulator},}\
  }\href@noop {} {\bibfield  {journal} {\bibinfo  {journal} {Europhys. Lett.}\
  }\textbf {\bibinfo {volume} {95}},\ \bibinfo {pages} {30006} (\bibinfo {year}
  {2011})}\BibitemShut {NoStop}%
\bibitem [{\citenamefont {Krutitsky}\ and\ \citenamefont
  {Navez}(2011)}]{Krutitsky:2011qf}%
  \BibitemOpen
  \bibfield  {author} {\bibinfo {author} {\bibfnamefont {Konstantin~V.}\
  \bibnamefont {Krutitsky}}\ and\ \bibinfo {author} {\bibfnamefont {Patrick}\
  \bibnamefont {Navez}},\ }\bibfield  {title} {\enquote {\bibinfo {title}
  {Excitation dynamics in a lattice bose gas within the time-dependent
  gutzwiller mean-field approach},}\ }\href {\doibase
  10.1103/PhysRevA.84.033602} {\bibfield  {journal} {\bibinfo  {journal} {Phys.
  Rev. A}\ }\textbf {\bibinfo {volume} {84}},\ \bibinfo {pages} {033602}
  (\bibinfo {year} {2011})}\BibitemShut {NoStop}%
\bibitem [{\citenamefont {Aoki}\ \emph {et~al.}(2014)\citenamefont {Aoki},
  \citenamefont {Tsuji}, \citenamefont {Eckstein}, \citenamefont {Kollar},
  \citenamefont {Oka},\ and\ \citenamefont {Werner}}]{Aoki:2014kx}%
  \BibitemOpen
  \bibfield  {author} {\bibinfo {author} {\bibfnamefont {Hideo}\ \bibnamefont
  {Aoki}}, \bibinfo {author} {\bibfnamefont {Naoto}\ \bibnamefont {Tsuji}},
  \bibinfo {author} {\bibfnamefont {Martin}\ \bibnamefont {Eckstein}}, \bibinfo
  {author} {\bibfnamefont {Marcus}\ \bibnamefont {Kollar}}, \bibinfo {author}
  {\bibfnamefont {Takashi}\ \bibnamefont {Oka}}, \ and\ \bibinfo {author}
  {\bibfnamefont {Philipp}\ \bibnamefont {Werner}},\ }\bibfield  {title}
  {\enquote {\bibinfo {title} {Nonequilibrium dynamical mean-field theory and
  its applications},}\ }\href {\doibase 10.1103/RevModPhys.86.779} {\bibfield
  {journal} {\bibinfo  {journal} {Rev. Mod. Phys.}\ }\textbf {\bibinfo {volume}
  {86}},\ \bibinfo {pages} {779--837} (\bibinfo {year} {2014})}\BibitemShut
  {NoStop}%
\bibitem [{\citenamefont {Byczuk}\ and\ \citenamefont
  {Vollhardt}(2008)}]{Byczuk:2008nx}%
  \BibitemOpen
  \bibfield  {author} {\bibinfo {author} {\bibfnamefont {Krzysztof}\
  \bibnamefont {Byczuk}}\ and\ \bibinfo {author} {\bibfnamefont {Dieter}\
  \bibnamefont {Vollhardt}},\ }\bibfield  {title} {\enquote {\bibinfo {title}
  {Correlated bosons on a lattice: Dynamical mean-field theory for
  bose-einstein condensed and normal phases},}\ }\href {\doibase
  10.1103/PhysRevB.77.235106} {\bibfield  {journal} {\bibinfo  {journal} {Phys.
  Rev. B}\ }\textbf {\bibinfo {volume} {77}},\ \bibinfo {pages} {235106}
  (\bibinfo {year} {2008})}\BibitemShut {NoStop}%
\bibitem [{\citenamefont {Hubener}\ \emph {et~al.}(2009)\citenamefont
  {Hubener}, \citenamefont {Snoek},\ and\ \citenamefont
  {Hofstetter}}]{Hubener:2009cr}%
  \BibitemOpen
  \bibfield  {author} {\bibinfo {author} {\bibfnamefont {A.}~\bibnamefont
  {Hubener}}, \bibinfo {author} {\bibfnamefont {M.}~\bibnamefont {Snoek}}, \
  and\ \bibinfo {author} {\bibfnamefont {W.}~\bibnamefont {Hofstetter}},\
  }\bibfield  {title} {\enquote {\bibinfo {title} {Magnetic phases of
  two-component ultracold bosons in an optical lattice},}\ }\href {\doibase
  10.1103/PhysRevB.80.245109} {\bibfield  {journal} {\bibinfo  {journal} {Phys.
  Rev. B}\ }\textbf {\bibinfo {volume} {80}},\ \bibinfo {pages} {245109}
  (\bibinfo {year} {2009})}\BibitemShut {NoStop}%
\bibitem [{\citenamefont {Anders}\ \emph {et~al.}(2010)\citenamefont {Anders},
  \citenamefont {Gull}, \citenamefont {Pollet}, \citenamefont {Troyer},\ and\
  \citenamefont {Werner}}]{Anders:2010uq}%
  \BibitemOpen
  \bibfield  {author} {\bibinfo {author} {\bibfnamefont {Peter}\ \bibnamefont
  {Anders}}, \bibinfo {author} {\bibfnamefont {Emanuel}\ \bibnamefont {Gull}},
  \bibinfo {author} {\bibfnamefont {Lode}\ \bibnamefont {Pollet}}, \bibinfo
  {author} {\bibfnamefont {Matthias}\ \bibnamefont {Troyer}}, \ and\ \bibinfo
  {author} {\bibfnamefont {Philipp}\ \bibnamefont {Werner}},\ }\bibfield
  {title} {\enquote {\bibinfo {title} {Dynamical mean field solution of the
  bose-hubbard model},}\ }\href {\doibase 10.1103/PhysRevLett.105.096402}
  {\bibfield  {journal} {\bibinfo  {journal} {Phys. Rev. Lett.}\ }\textbf
  {\bibinfo {volume} {105}},\ \bibinfo {pages} {096402} (\bibinfo {year}
  {2010})}\BibitemShut {NoStop}%
\bibitem [{\citenamefont {Anders}\ \emph {et~al.}(2011)\citenamefont {Anders},
  \citenamefont {Gull}, \citenamefont {Pollet}, \citenamefont {Troyer},\ and\
  \citenamefont {Werner}}]{Anders:2011uq}%
  \BibitemOpen
  \bibfield  {author} {\bibinfo {author} {\bibfnamefont {Peter}\ \bibnamefont
  {Anders}}, \bibinfo {author} {\bibfnamefont {Emanuel}\ \bibnamefont {Gull}},
  \bibinfo {author} {\bibfnamefont {Lode}\ \bibnamefont {Pollet}}, \bibinfo
  {author} {\bibfnamefont {Matthias}\ \bibnamefont {Troyer}}, \ and\ \bibinfo
  {author} {\bibfnamefont {Philipp}\ \bibnamefont {Werner}},\ }\bibfield
  {title} {\enquote {\bibinfo {title} {Dynamical mean-field theory for
  bosons},}\ }\href@noop {} {\bibfield  {journal} {\bibinfo  {journal} {New J.
  Phys.}\ }\textbf {\bibinfo {volume} {13}},\ \bibinfo {pages} {075013}
  (\bibinfo {year} {2011})}\BibitemShut {NoStop}%
\bibitem [{\citenamefont {Werner}\ \emph {et~al.}(2009)\citenamefont {Werner},
  \citenamefont {Oka},\ and\ \citenamefont {Millis}}]{Werner:2009tg}%
  \BibitemOpen
  \bibfield  {author} {\bibinfo {author} {\bibfnamefont {Philipp}\ \bibnamefont
  {Werner}}, \bibinfo {author} {\bibfnamefont {Takashi}\ \bibnamefont {Oka}}, \
  and\ \bibinfo {author} {\bibfnamefont {Andrew~J.}\ \bibnamefont {Millis}},\
  }\bibfield  {title} {\enquote {\bibinfo {title} {Diagrammatic monte carlo
  simulation of nonequilibrium systems},}\ }\href {\doibase
  10.1103/PhysRevB.79.035320} {\bibfield  {journal} {\bibinfo  {journal} {Phys.
  Rev. B}\ }\textbf {\bibinfo {volume} {79}},\ \bibinfo {pages} {035320}
  (\bibinfo {year} {2009})}\BibitemShut {NoStop}%
\bibitem [{\citenamefont {Gramsch}\ \emph {et~al.}(2013)\citenamefont
  {Gramsch}, \citenamefont {Balzer}, \citenamefont {Eckstein},\ and\
  \citenamefont {Kollar}}]{Gramsch:2013fk}%
  \BibitemOpen
  \bibfield  {author} {\bibinfo {author} {\bibfnamefont {Christian}\
  \bibnamefont {Gramsch}}, \bibinfo {author} {\bibfnamefont {Karsten}\
  \bibnamefont {Balzer}}, \bibinfo {author} {\bibfnamefont {Martin}\
  \bibnamefont {Eckstein}}, \ and\ \bibinfo {author} {\bibfnamefont {Marcus}\
  \bibnamefont {Kollar}},\ }\bibfield  {title} {\enquote {\bibinfo {title}
  {Hamiltonian-based impurity solver for nonequilibrium dynamical mean-field
  theory},}\ }\href {\doibase 10.1103/PhysRevB.88.235106} {\bibfield  {journal}
  {\bibinfo  {journal} {Phys. Rev. B}\ }\textbf {\bibinfo {volume} {88}},\
  \bibinfo {pages} {235106} (\bibinfo {year} {2013})}\BibitemShut {NoStop}%
\bibitem [{\citenamefont {Keiter}\ and\ \citenamefont
  {Kimball}(1971)}]{Keiter:1971hc}%
  \BibitemOpen
  \bibfield  {author} {\bibinfo {author} {\bibfnamefont {H.}~\bibnamefont
  {Keiter}}\ and\ \bibinfo {author} {\bibfnamefont {J.~C.}\ \bibnamefont
  {Kimball}},\ }\bibfield  {title} {\enquote {\bibinfo {title} {Diagrammatic
  approach to the anderson model for dilute alloys},}\ }\href@noop {}
  {\bibfield  {journal} {\bibinfo  {journal} {J. Appl. Phys.}\ }\textbf
  {\bibinfo {volume} {42}},\ \bibinfo {pages} {1460--1461} (\bibinfo {year}
  {1971})}\BibitemShut {NoStop}%
\bibitem [{\citenamefont {Stefanucci}\ and\ \citenamefont {van
  Leeuwen}(2013)}]{Stefanucci:2013oq}%
  \BibitemOpen
  \bibfield  {author} {\bibinfo {author} {\bibfnamefont {Gianluca}\
  \bibnamefont {Stefanucci}}\ and\ \bibinfo {author} {\bibfnamefont {Robert}\
  \bibnamefont {van Leeuwen}},\ }\href@noop {} {\emph {\bibinfo {title}
  {Nonequilibrium Many-Body Theory of Quantum Systems A Modern Introduction}}}\
  (\bibinfo  {publisher} {Cambridge University Press},\ \bibinfo {year}
  {2013})\BibitemShut {NoStop}%
\bibitem [{\citenamefont {Sachdev}(1999)}]{Sachdev:1999fk}%
  \BibitemOpen
  \bibfield  {author} {\bibinfo {author} {\bibfnamefont {Subir}\ \bibnamefont
  {Sachdev}},\ }\href@noop {} {\emph {\bibinfo {title} {Quantum Phase
  Transitions}}}\ (\bibinfo  {publisher} {Cambridge University Press},\
  \bibinfo {address} {The Edinburgh Building, Cambridge CB2 2RU, UK},\ \bibinfo
  {year} {1999})\BibitemShut {NoStop}%
\bibitem [{\citenamefont {Werner}\ \emph {et~al.}(2010)\citenamefont {Werner},
  \citenamefont {Oka}, \citenamefont {Eckstein},\ and\ \citenamefont
  {Millis}}]{Werner:2010pb}%
  \BibitemOpen
  \bibfield  {author} {\bibinfo {author} {\bibfnamefont {Philipp}\ \bibnamefont
  {Werner}}, \bibinfo {author} {\bibfnamefont {Takashi}\ \bibnamefont {Oka}},
  \bibinfo {author} {\bibfnamefont {Martin}\ \bibnamefont {Eckstein}}, \ and\
  \bibinfo {author} {\bibfnamefont {Andrew~J.}\ \bibnamefont {Millis}},\
  }\bibfield  {title} {\enquote {\bibinfo {title} {Weak-coupling quantum monte
  carlo calculations on the keldysh contour: Theory and application to the
  current-voltage characteristics of the anderson model},}\ }\href {\doibase
  10.1103/PhysRevB.81.035108} {\bibfield  {journal} {\bibinfo  {journal} {Phys.
  Rev. B}\ }\textbf {\bibinfo {volume} {81}},\ \bibinfo {pages} {035108}
  (\bibinfo {year} {2010})}\BibitemShut {NoStop}%
\bibitem [{\citenamefont {Eckstein}\ and\ \citenamefont
  {Werner}(2010)}]{Eckstein:2010fk}%
  \BibitemOpen
  \bibfield  {author} {\bibinfo {author} {\bibfnamefont {Martin}\ \bibnamefont
  {Eckstein}}\ and\ \bibinfo {author} {\bibfnamefont {Philipp}\ \bibnamefont
  {Werner}},\ }\bibfield  {title} {\enquote {\bibinfo {title} {Nonequilibrium
  dynamical mean-field calculations based on the noncrossing approximation and
  its generalizations},}\ }\href {\doibase 10.1103/PhysRevB.82.115115}
  {\bibfield  {journal} {\bibinfo  {journal} {Phys. Rev. B}\ }\textbf {\bibinfo
  {volume} {82}},\ \bibinfo {pages} {115115} (\bibinfo {year}
  {2010})}\BibitemShut {NoStop}%
\bibitem [{\citenamefont {Brunner}\ and\ \citenamefont {van~der
  Houwen}(1986)}]{Brunner:1986ff}%
  \BibitemOpen
  \bibfield  {author} {\bibinfo {author} {\bibfnamefont {H.}~\bibnamefont
  {Brunner}}\ and\ \bibinfo {author} {\bibfnamefont {P.~J.}\ \bibnamefont
  {van~der Houwen}},\ }\href@noop {} {\emph {\bibinfo {title} {The Numerical
  Solution of Volterra Equations}}}\ (\bibinfo  {publisher} {North-Holland},\
  \bibinfo {address} {Amsterdam},\ \bibinfo {year} {1986})\BibitemShut
  {NoStop}%
\bibitem [{\citenamefont {Teichmann}\ \emph {et~al.}(2009)\citenamefont
  {Teichmann}, \citenamefont {Hinrichs}, \citenamefont {Holthaus},\ and\
  \citenamefont {Eckardt}}]{Teichmann:2009bh}%
  \BibitemOpen
  \bibfield  {author} {\bibinfo {author} {\bibfnamefont {Niklas}\ \bibnamefont
  {Teichmann}}, \bibinfo {author} {\bibfnamefont {Dennis}\ \bibnamefont
  {Hinrichs}}, \bibinfo {author} {\bibfnamefont {Martin}\ \bibnamefont
  {Holthaus}}, \ and\ \bibinfo {author} {\bibfnamefont {Andr\'e}\ \bibnamefont
  {Eckardt}},\ }\bibfield  {title} {\enquote {\bibinfo {title} {Process-chain
  approach to the bose-hubbard model: Ground-state properties and phase
  diagram},}\ }\href {\doibase 10.1103/PhysRevB.79.224515} {\bibfield
  {journal} {\bibinfo  {journal} {Phys. Rev. B}\ }\textbf {\bibinfo {volume}
  {79}},\ \bibinfo {pages} {224515} (\bibinfo {year} {2009})}\BibitemShut
  {NoStop}%
\bibitem [{\citenamefont {Kauch}\ \emph {et~al.}(2012)\citenamefont {Kauch},
  \citenamefont {Byczuk},\ and\ \citenamefont {Vollhardt}}]{Kauch:2012ve}%
  \BibitemOpen
  \bibfield  {author} {\bibinfo {author} {\bibfnamefont {Anna}\ \bibnamefont
  {Kauch}}, \bibinfo {author} {\bibfnamefont {Krzysztof}\ \bibnamefont
  {Byczuk}}, \ and\ \bibinfo {author} {\bibfnamefont {Dieter}\ \bibnamefont
  {Vollhardt}},\ }\bibfield  {title} {\enquote {\bibinfo {title}
  {Strong-coupling solution of the bosonic dynamical mean-field theory},}\
  }\href {\doibase 10.1103/PhysRevB.85.205115} {\bibfield  {journal} {\bibinfo
  {journal} {Phys. Rev. B}\ }\textbf {\bibinfo {volume} {85}},\ \bibinfo
  {pages} {205115} (\bibinfo {year} {2012})}\BibitemShut {NoStop}%
\bibitem [{\citenamefont {Pruschke}\ \emph {et~al.}(1993)\citenamefont
  {Pruschke}, \citenamefont {Cox},\ and\ \citenamefont
  {Jarrell}}]{Pruschke:1993aa}%
  \BibitemOpen
  \bibfield  {author} {\bibinfo {author} {\bibfnamefont {Th.}\ \bibnamefont
  {Pruschke}}, \bibinfo {author} {\bibfnamefont {D.~L.}\ \bibnamefont {Cox}}, \
  and\ \bibinfo {author} {\bibfnamefont {M.}~\bibnamefont {Jarrell}},\
  }\bibfield  {title} {\enquote {\bibinfo {title} {Hubbard model at infinite
  dimensions: Thermodynamic and transport properties},}\ }\href@noop {}
  {\bibfield  {journal} {\bibinfo  {journal} {Phys. Rev. B}\ }\textbf {\bibinfo
  {volume} {47}},\ \bibinfo {pages} {3553} (\bibinfo {year}
  {1993})}\BibitemShut {NoStop}%
\bibitem [{\citenamefont {Freericks}\ \emph {et~al.}(2013)\citenamefont
  {Freericks}, \citenamefont {Turkowski}, \citenamefont {Krishnamurthy},\ and\
  \citenamefont {Knap}}]{Freericks:2013oq}%
  \BibitemOpen
  \bibfield  {author} {\bibinfo {author} {\bibfnamefont {J.~K.}\ \bibnamefont
  {Freericks}}, \bibinfo {author} {\bibfnamefont {V.}~\bibnamefont
  {Turkowski}}, \bibinfo {author} {\bibfnamefont {H.~R.}\ \bibnamefont
  {Krishnamurthy}}, \ and\ \bibinfo {author} {\bibfnamefont {M.}~\bibnamefont
  {Knap}},\ }\bibfield  {title} {\enquote {\bibinfo {title} {Spectral moment
  sum rules for the retarded green's function and self-energy of the
  inhomogeneous bose-hubbard model in equilibrium and nonequilibrium},}\ }\href
  {\doibase 10.1103/PhysRevA.87.013628} {\bibfield  {journal} {\bibinfo
  {journal} {Phys. Rev. A}\ }\textbf {\bibinfo {volume} {87}},\ \bibinfo
  {pages} {013628} (\bibinfo {year} {2013})}\BibitemShut {NoStop}%
\bibitem [{\citenamefont {Eckstein}\ \emph {et~al.}(2009)\citenamefont
  {Eckstein}, \citenamefont {Kollar},\ and\ \citenamefont
  {Werner}}]{Eckstein:2009fu}%
  \BibitemOpen
  \bibfield  {author} {\bibinfo {author} {\bibfnamefont {Martin}\ \bibnamefont
  {Eckstein}}, \bibinfo {author} {\bibfnamefont {Marcus}\ \bibnamefont
  {Kollar}}, \ and\ \bibinfo {author} {\bibfnamefont {Philipp}\ \bibnamefont
  {Werner}},\ }\bibfield  {title} {\enquote {\bibinfo {title} {Thermalization
  after an interaction quench in the hubbard model},}\ }\href {\doibase
  10.1103/PhysRevLett.103.056403} {\bibfield  {journal} {\bibinfo  {journal}
  {Phys. Rev. Lett.}\ }\textbf {\bibinfo {volume} {103}},\ \bibinfo {pages}
  {056403} (\bibinfo {year} {2009})}\BibitemShut {NoStop}%
\bibitem [{\citenamefont {Berges}\ and\ \citenamefont
  {Sexty}(2012)}]{Berges:2012uq}%
  \BibitemOpen
  \bibfield  {author} {\bibinfo {author} {\bibfnamefont {J\"urgen}\
  \bibnamefont {Berges}}\ and\ \bibinfo {author} {\bibfnamefont {D\'enes}\
  \bibnamefont {Sexty}},\ }\bibfield  {title} {\enquote {\bibinfo {title}
  {Bose-einstein condensation in relativistic field theories far from
  equilibrium},}\ }\href {\doibase 10.1103/PhysRevLett.108.161601} {\bibfield
  {journal} {\bibinfo  {journal} {Phys. Rev. Lett.}\ }\textbf {\bibinfo
  {volume} {108}},\ \bibinfo {pages} {161601} (\bibinfo {year}
  {2012})}\BibitemShut {NoStop}%
\bibitem [{\citenamefont {Machholm}\ \emph {et~al.}(2003)\citenamefont
  {Machholm}, \citenamefont {Pethick},\ and\ \citenamefont
  {Smith}}]{Machholm:2003ly}%
  \BibitemOpen
  \bibfield  {author} {\bibinfo {author} {\bibfnamefont {M.}~\bibnamefont
  {Machholm}}, \bibinfo {author} {\bibfnamefont {C.~J.}\ \bibnamefont
  {Pethick}}, \ and\ \bibinfo {author} {\bibfnamefont {H.}~\bibnamefont
  {Smith}},\ }\bibfield  {title} {\enquote {\bibinfo {title} {Band structure,
  elementary excitations, and stability of a bose-einstein condensate in a
  periodic potential},}\ }\href {\doibase 10.1103/PhysRevA.67.053613}
  {\bibfield  {journal} {\bibinfo  {journal} {Phys. Rev. A}\ }\textbf {\bibinfo
  {volume} {67}},\ \bibinfo {pages} {053613} (\bibinfo {year}
  {2003})}\BibitemShut {NoStop}%
\bibitem [{\citenamefont {De~Sarlo}\ \emph {et~al.}(2005)\citenamefont
  {De~Sarlo}, \citenamefont {Fallani}, \citenamefont {Lye}, \citenamefont
  {Modugno}, \citenamefont {Saers}, \citenamefont {Fort},\ and\ \citenamefont
  {Inguscio}}]{De-Sarlo:2005kl}%
  \BibitemOpen
  \bibfield  {author} {\bibinfo {author} {\bibfnamefont {L.}~\bibnamefont
  {De~Sarlo}}, \bibinfo {author} {\bibfnamefont {L.}~\bibnamefont {Fallani}},
  \bibinfo {author} {\bibfnamefont {J.~E.}\ \bibnamefont {Lye}}, \bibinfo
  {author} {\bibfnamefont {M.}~\bibnamefont {Modugno}}, \bibinfo {author}
  {\bibfnamefont {R.}~\bibnamefont {Saers}}, \bibinfo {author} {\bibfnamefont
  {C.}~\bibnamefont {Fort}}, \ and\ \bibinfo {author} {\bibfnamefont
  {M.}~\bibnamefont {Inguscio}},\ }\bibfield  {title} {\enquote {\bibinfo
  {title} {Unstable regimes for a bose-einstein condensate in an optical
  lattice},}\ }\href {\doibase 10.1103/PhysRevA.72.013603} {\bibfield
  {journal} {\bibinfo  {journal} {Phys. Rev. A}\ }\textbf {\bibinfo {volume}
  {72}},\ \bibinfo {pages} {013603} (\bibinfo {year} {2005})}\BibitemShut
  {NoStop}%
\bibitem [{\citenamefont {Fischer}\ and\ \citenamefont
  {Sch\"utzhold}(2008)}]{Fischer:2008ys}%
  \BibitemOpen
  \bibfield  {author} {\bibinfo {author} {\bibfnamefont {Uwe~R.}\ \bibnamefont
  {Fischer}}\ and\ \bibinfo {author} {\bibfnamefont {Ralf}\ \bibnamefont
  {Sch\"utzhold}},\ }\bibfield  {title} {\enquote {\bibinfo {title}
  {Tunneling-induced damping of phase coherence revivals in deep optical
  lattices},}\ }\href {\doibase 10.1103/PhysRevA.78.061603} {\bibfield
  {journal} {\bibinfo  {journal} {Phys. Rev. A}\ }\textbf {\bibinfo {volume}
  {78}},\ \bibinfo {pages} {061603} (\bibinfo {year} {2008})}\BibitemShut
  {NoStop}%
\bibitem [{Note1()}]{Note1}%
  \BibitemOpen
  \bibinfo {note} {Note that the $N_{\protect \textrm {max}}> 3$ results of
  Refs.\ \protect \rev@citealp {Sciolla:2010uq} and \protect \rev@citealp
  {Sciolla:2011kx} are in the high-$U_i$ regime.}\BibitemShut {Stop}%
\bibitem [{\citenamefont {Schir\'o}\ and\ \citenamefont
  {Fabrizio}(2010)}]{Schiro:2010fk}%
  \BibitemOpen
  \bibfield  {author} {\bibinfo {author} {\bibfnamefont {Marco}\ \bibnamefont
  {Schir\'o}}\ and\ \bibinfo {author} {\bibfnamefont {Michele}\ \bibnamefont
  {Fabrizio}},\ }\bibfield  {title} {\enquote {\bibinfo {title} {Time-dependent
  mean field theory for quench dynamics in correlated electron systems},}\
  }\href {\doibase 10.1103/PhysRevLett.105.076401} {\bibfield  {journal}
  {\bibinfo  {journal} {Phys. Rev. Lett.}\ }\textbf {\bibinfo {volume} {105}},\
  \bibinfo {pages} {076401} (\bibinfo {year} {2010})}\BibitemShut {NoStop}%
\bibitem [{\citenamefont {Gambassi}\ and\ \citenamefont
  {Calabrese}(2011)}]{Gambassi:2011dz}%
  \BibitemOpen
  \bibfield  {author} {\bibinfo {author} {\bibfnamefont {A.}~\bibnamefont
  {Gambassi}}\ and\ \bibinfo {author} {\bibfnamefont {P.}~\bibnamefont
  {Calabrese}},\ }\bibfield  {title} {\enquote {\bibinfo {title} {Quantum
  quenches as classical critical films},}\ }\href@noop {} {\bibfield  {journal}
  {\bibinfo  {journal} {Europhys. Lett.}\ }\textbf {\bibinfo {volume} {95}},\
  \bibinfo {pages} {66007} (\bibinfo {year} {2011})}\BibitemShut {NoStop}%
\bibitem [{\citenamefont {Schir\'o}\ and\ \citenamefont
  {Fabrizio}(2011)}]{Schiro:2011oq}%
  \BibitemOpen
  \bibfield  {author} {\bibinfo {author} {\bibfnamefont {Marco}\ \bibnamefont
  {Schir\'o}}\ and\ \bibinfo {author} {\bibfnamefont {Michele}\ \bibnamefont
  {Fabrizio}},\ }\bibfield  {title} {\enquote {\bibinfo {title} {Quantum
  quenches in the hubbard model: Time-dependent mean-field theory and the role
  of quantum fluctuations},}\ }\href {\doibase 10.1103/PhysRevB.83.165105}
  {\bibfield  {journal} {\bibinfo  {journal} {Phys. Rev. B}\ }\textbf {\bibinfo
  {volume} {83}},\ \bibinfo {pages} {165105} (\bibinfo {year}
  {2011})}\BibitemShut {NoStop}%
\bibitem [{\citenamefont {Sandri}\ \emph {et~al.}(2012)\citenamefont {Sandri},
  \citenamefont {Schir\'o},\ and\ \citenamefont {Fabrizio}}]{Sandri:2012fv}%
  \BibitemOpen
  \bibfield  {author} {\bibinfo {author} {\bibfnamefont {Matteo}\ \bibnamefont
  {Sandri}}, \bibinfo {author} {\bibfnamefont {Marco}\ \bibnamefont
  {Schir\'o}}, \ and\ \bibinfo {author} {\bibfnamefont {Michele}\ \bibnamefont
  {Fabrizio}},\ }\bibfield  {title} {\enquote {\bibinfo {title} {Linear ramps
  of interaction in the fermionic hubbard model},}\ }\href {\doibase
  10.1103/PhysRevB.86.075122} {\bibfield  {journal} {\bibinfo  {journal} {Phys.
  Rev. B}\ }\textbf {\bibinfo {volume} {86}},\ \bibinfo {pages} {075122}
  (\bibinfo {year} {2012})}\BibitemShut {NoStop}%
\bibitem [{\citenamefont {Sciolla}\ and\ \citenamefont
  {Biroli}(2013)}]{Sciolla:2013bs}%
  \BibitemOpen
  \bibfield  {author} {\bibinfo {author} {\bibfnamefont {Bruno}\ \bibnamefont
  {Sciolla}}\ and\ \bibinfo {author} {\bibfnamefont {Giulio}\ \bibnamefont
  {Biroli}},\ }\bibfield  {title} {\enquote {\bibinfo {title} {Quantum
  quenches, dynamical transitions, and off-equilibrium quantum criticality},}\
  }\href {\doibase 10.1103/PhysRevB.88.201110} {\bibfield  {journal} {\bibinfo
  {journal} {Phys. Rev. B}\ }\textbf {\bibinfo {volume} {88}},\ \bibinfo
  {pages} {201110} (\bibinfo {year} {2013})}\BibitemShut {NoStop}%
\bibitem [{\citenamefont {Schneider}(2014)}]{Schneider:2014ve}%
  \BibitemOpen
  \bibfield  {author} {\bibinfo {author} {\bibfnamefont {Ulrich}\ \bibnamefont
  {Schneider}},\ }\href@noop {} {}\bibinfo {howpublished} {private
  communication} (\bibinfo {year} {2014})\BibitemShut {NoStop}%
\bibitem [{\citenamefont {Altman}\ and\ \citenamefont
  {Auerbach}(2002)}]{Altman:2002ve}%
  \BibitemOpen
  \bibfield  {author} {\bibinfo {author} {\bibfnamefont {Ehud}\ \bibnamefont
  {Altman}}\ and\ \bibinfo {author} {\bibfnamefont {Assa}\ \bibnamefont
  {Auerbach}},\ }\bibfield  {title} {\enquote {\bibinfo {title} {Oscillating
  superfluidity of bosons in optical lattices},}\ }\href {\doibase
  10.1103/PhysRevLett.89.250404} {\bibfield  {journal} {\bibinfo  {journal}
  {Phys. Rev. Lett.}\ }\textbf {\bibinfo {volume} {89}},\ \bibinfo {pages}
  {250404} (\bibinfo {year} {2002})}\BibitemShut {NoStop}%
\bibitem [{\citenamefont {S{\"o}yler}\ \emph {et~al.}(2009)\citenamefont
  {S{\"o}yler}, \citenamefont {Capogrosso-Sansone}, \citenamefont {Prokof'ev},\
  and\ \citenamefont {Svistunov}}]{Soyler:2009ys}%
  \BibitemOpen
  \bibfield  {author} {\bibinfo {author} {\bibfnamefont {{\c S}~G}\
  \bibnamefont {S{\"o}yler}}, \bibinfo {author} {\bibfnamefont {B}~\bibnamefont
  {Capogrosso-Sansone}}, \bibinfo {author} {\bibfnamefont {N~V}\ \bibnamefont
  {Prokof'ev}}, \ and\ \bibinfo {author} {\bibfnamefont {B~V}\ \bibnamefont
  {Svistunov}},\ }\bibfield  {title} {\enquote {\bibinfo {title}
  {Sign-alternating interaction mediated by strongly correlated lattice
  bosons},}\ }\href@noop {} {\bibfield  {journal} {\bibinfo  {journal} {New J.
  Phys.}\ }\textbf {\bibinfo {volume} {11}},\ \bibinfo {pages} {073036}
  (\bibinfo {year} {2009})}\BibitemShut {NoStop}%
\bibitem [{\citenamefont {Capogrosso-Sansone}\ \emph
  {et~al.}(2010{\natexlab{a}})\citenamefont {Capogrosso-Sansone}, \citenamefont
  {S{\"o}yler}, \citenamefont {Prokof'ev},\ and\ \citenamefont
  {Svistunov}}]{Capogrosso-Sansone:2010zr}%
  \BibitemOpen
  \bibfield  {author} {\bibinfo {author} {\bibfnamefont {B.}~\bibnamefont
  {Capogrosso-Sansone}}, \bibinfo {author} {\bibfnamefont {{\c S}.~G.}\
  \bibnamefont {S{\"o}yler}}, \bibinfo {author} {\bibfnamefont {N.~V.}\
  \bibnamefont {Prokof'ev}}, \ and\ \bibinfo {author} {\bibfnamefont {B.~V.}\
  \bibnamefont {Svistunov}},\ }\bibfield  {title} {\enquote {\bibinfo {title}
  {Critical entropies for magnetic ordering in bosonic mixtures on a
  lattice},}\ }\href {\doibase 10.1103/PhysRevA.81.053622} {\bibfield
  {journal} {\bibinfo  {journal} {Phys. Rev. A}\ }\textbf {\bibinfo {volume}
  {81}},\ \bibinfo {pages} {053622} (\bibinfo {year}
  {2010}{\natexlab{a}})}\BibitemShut {NoStop}%
\bibitem [{\citenamefont {Eckstein}\ and\ \citenamefont
  {Werner}(2013)}]{Eckstein:2013fv}%
  \BibitemOpen
  \bibfield  {author} {\bibinfo {author} {\bibfnamefont {Martin}\ \bibnamefont
  {Eckstein}}\ and\ \bibinfo {author} {\bibfnamefont {Philipp}\ \bibnamefont
  {Werner}},\ }\bibfield  {title} {\enquote {\bibinfo {title} {Nonequilibrium
  dynamical mean-field simulation of inhomogeneous systems},}\ }\href {\doibase
  10.1103/PhysRevB.88.075135} {\bibfield  {journal} {\bibinfo  {journal} {Phys.
  Rev. B}\ }\textbf {\bibinfo {volume} {88}},\ \bibinfo {pages} {075135}
  (\bibinfo {year} {2013})}\BibitemShut {NoStop}%
\bibitem [{\citenamefont {Johnson}\ \emph {et~al.}(2009)\citenamefont
  {Johnson}, \citenamefont {Tiesinga}, \citenamefont {Porto},\ and\
  \citenamefont {Williams}}]{Johnson:2009kx}%
  \BibitemOpen
  \bibfield  {author} {\bibinfo {author} {\bibfnamefont {P~R}\ \bibnamefont
  {Johnson}}, \bibinfo {author} {\bibfnamefont {E}~\bibnamefont {Tiesinga}},
  \bibinfo {author} {\bibfnamefont {J~V}\ \bibnamefont {Porto}}, \ and\
  \bibinfo {author} {\bibfnamefont {C~J}\ \bibnamefont {Williams}},\ }\bibfield
   {title} {\enquote {\bibinfo {title} {Effective three-body interactions of
  neutral bosons in optical lattices},}\ }\href@noop {} {\bibfield  {journal}
  {\bibinfo  {journal} {New J. Phys.}\ }\textbf {\bibinfo {volume} {11}},\
  \bibinfo {pages} {093022} (\bibinfo {year} {2009})}\BibitemShut {NoStop}%
\bibitem [{\citenamefont {Tiesinga}\ and\ \citenamefont
  {Johnson}(2011)}]{Tiesinga:2011ve}%
  \BibitemOpen
  \bibfield  {author} {\bibinfo {author} {\bibfnamefont {E.}~\bibnamefont
  {Tiesinga}}\ and\ \bibinfo {author} {\bibfnamefont {P.~R.}\ \bibnamefont
  {Johnson}},\ }\bibfield  {title} {\enquote {\bibinfo {title} {Collapse and
  revival dynamics of number-squeezed superfluids of ultracold atoms in optical
  lattices},}\ }\href {\doibase 10.1103/PhysRevA.83.063609} {\bibfield
  {journal} {\bibinfo  {journal} {Phys. Rev. A}\ }\textbf {\bibinfo {volume}
  {83}},\ \bibinfo {pages} {063609} (\bibinfo {year} {2011})}\BibitemShut
  {NoStop}%
\bibitem [{\citenamefont {Dirks}\ \emph {et~al.}(2014)\citenamefont {Dirks},
  \citenamefont {Mikelsons}, \citenamefont {Krishnamurthy},\ and\ \citenamefont
  {Freericks}}]{Dirks:2014ij}%
  \BibitemOpen
  \bibfield  {author} {\bibinfo {author} {\bibfnamefont {Andreas}\ \bibnamefont
  {Dirks}}, \bibinfo {author} {\bibfnamefont {Karlis}\ \bibnamefont
  {Mikelsons}}, \bibinfo {author} {\bibfnamefont {H.~R.}\ \bibnamefont
  {Krishnamurthy}}, \ and\ \bibinfo {author} {\bibfnamefont {James~K.}\
  \bibnamefont {Freericks}},\ }\bibfield  {title} {\enquote {\bibinfo {title}
  {Simulation of inhomogeneous distributions of ultracold atoms in an optical
  lattice via a massively parallel implementation of nonequilibrium
  strong-coupling perturbation theory},}\ }\href {\doibase
  10.1103/PhysRevE.89.023306} {\bibfield  {journal} {\bibinfo  {journal} {Phys.
  Rev. E}\ }\textbf {\bibinfo {volume} {89}},\ \bibinfo {pages} {023306}
  (\bibinfo {year} {2014})}\BibitemShut {NoStop}%
\bibitem [{\citenamefont {Carusotto}\ and\ \citenamefont
  {Ciuti}(2013)}]{Carusotto:2013kx}%
  \BibitemOpen
  \bibfield  {author} {\bibinfo {author} {\bibfnamefont {Iacopo}\ \bibnamefont
  {Carusotto}}\ and\ \bibinfo {author} {\bibfnamefont {Cristiano}\ \bibnamefont
  {Ciuti}},\ }\bibfield  {title} {\enquote {\bibinfo {title} {Quantum fluids of
  light},}\ }\href {\doibase 10.1103/RevModPhys.85.299} {\bibfield  {journal}
  {\bibinfo  {journal} {Rev. Mod. Phys.}\ }\textbf {\bibinfo {volume} {85}},\
  \bibinfo {pages} {299--366} (\bibinfo {year} {2013})}\BibitemShut {NoStop}%
\bibitem [{\citenamefont {Akerlund}\ \emph {et~al.}(2013)\citenamefont
  {Akerlund}, \citenamefont {de~Forcrand}, \citenamefont {Georges},\ and\
  \citenamefont {Werner}}]{Akerlund:2013fk}%
  \BibitemOpen
  \bibfield  {author} {\bibinfo {author} {\bibfnamefont {Oscar}\ \bibnamefont
  {Akerlund}}, \bibinfo {author} {\bibfnamefont {Philippe}\ \bibnamefont
  {de~Forcrand}}, \bibinfo {author} {\bibfnamefont {Antoine}\ \bibnamefont
  {Georges}}, \ and\ \bibinfo {author} {\bibfnamefont {Philipp}\ \bibnamefont
  {Werner}},\ }\bibfield  {title} {\enquote {\bibinfo {title} {Dynamical mean
  field approximation applied to quantum field theory},}\ }\href {\doibase
  10.1103/PhysRevD.88.125006} {\bibfield  {journal} {\bibinfo  {journal} {Phys.
  Rev. D}\ }\textbf {\bibinfo {volume} {88}},\ \bibinfo {pages} {125006}
  (\bibinfo {year} {2013})}\BibitemShut {NoStop}%
\bibitem [{\citenamefont {{Snoek}}\ and\ \citenamefont
  {{Hofstetter}}(2010)}]{Snoek:2010uq}%
  \BibitemOpen
  \bibfield  {author} {\bibinfo {author} {\bibfnamefont {M.}~\bibnamefont
  {{Snoek}}}\ and\ \bibinfo {author} {\bibfnamefont {W.}~\bibnamefont
  {{Hofstetter}}},\ }\bibfield  {title} {\enquote {\bibinfo {title} {{Bosonic
  Dynamical Mean-Field Theory}},}\ }\href@noop {} {\bibfield  {journal}
  {\bibinfo  {journal} {ArXiv e-prints}\ } (\bibinfo {year} {2010})},\ \Eprint
  {http://arxiv.org/abs/1007.5223} {arXiv:1007.5223 [cond-mat.quant-gas]}
  \BibitemShut {NoStop}%
\bibitem [{\citenamefont {Georges}\ \emph {et~al.}(1996)\citenamefont
  {Georges}, \citenamefont {Kotliar}, \citenamefont {Krauth},\ and\
  \citenamefont {Rozenberg}}]{Georges:1996aa}%
  \BibitemOpen
  \bibfield  {author} {\bibinfo {author} {\bibfnamefont {Antoine}\ \bibnamefont
  {Georges}}, \bibinfo {author} {\bibfnamefont {Gabriel}\ \bibnamefont
  {Kotliar}}, \bibinfo {author} {\bibfnamefont {Werner}\ \bibnamefont
  {Krauth}}, \ and\ \bibinfo {author} {\bibfnamefont {Marcelo~J.}\ \bibnamefont
  {Rozenberg}},\ }\bibfield  {title} {\enquote {\bibinfo {title} {Dynamical
  mean-field theory of strongly correlated fermion systems and the limit of
  infinite dimensions},}\ }\href@noop {} {\bibfield  {journal} {\bibinfo
  {journal} {Rev. Mod. Phys.}\ }\textbf {\bibinfo {volume} {68}},\ \bibinfo
  {pages} {13--125} (\bibinfo {year} {1996})}\BibitemShut {NoStop}%
\bibitem [{\citenamefont {Kubo}(1962)}]{JPSJ.17.1100}%
  \BibitemOpen
  \bibfield  {author} {\bibinfo {author} {\bibfnamefont {Ryogo}\ \bibnamefont
  {Kubo}},\ }\bibfield  {title} {\enquote {\bibinfo {title} {Generalized
  cumulant expansion method},}\ }\href {\doibase 10.1143/JPSJ.17.1100}
  {\bibfield  {journal} {\bibinfo  {journal} {Journal of the Physical Society
  of Japan}\ }\textbf {\bibinfo {volume} {17}},\ \bibinfo {pages} {1100--1120}
  (\bibinfo {year} {1962})}\BibitemShut {NoStop}%
\bibitem [{\citenamefont {Anders}(2011)}]{Anders:2011fk}%
  \BibitemOpen
  \bibfield  {author} {\bibinfo {author} {\bibfnamefont {Peter~Christian}\
  \bibnamefont {Anders}},\ }\emph {\bibinfo {title} {Dynamical Mean-Field
  Theory for Bosons and Bose-Fermi Mixtures}},\ \href@noop {} {Ph.D. thesis},\
  \bibinfo  {school} {ETH Z{\"u}rich} (\bibinfo {year} {2011})\BibitemShut
  {NoStop}%
\bibitem [{\citenamefont {Capogrosso-Sansone}\ \emph
  {et~al.}(2010{\natexlab{b}})\citenamefont {Capogrosso-Sansone}, \citenamefont
  {Giorgini}, \citenamefont {Pilati}, \citenamefont {Pollet}, \citenamefont
  {Prokof'ev}, \citenamefont {Svistunov},\ and\ \citenamefont
  {Troyer}}]{Capogrosso-Sansone:2010vn}%
  \BibitemOpen
  \bibfield  {author} {\bibinfo {author} {\bibfnamefont {B}~\bibnamefont
  {Capogrosso-Sansone}}, \bibinfo {author} {\bibfnamefont {S}~\bibnamefont
  {Giorgini}}, \bibinfo {author} {\bibfnamefont {S}~\bibnamefont {Pilati}},
  \bibinfo {author} {\bibfnamefont {L}~\bibnamefont {Pollet}}, \bibinfo
  {author} {\bibfnamefont {N}~\bibnamefont {Prokof'ev}}, \bibinfo {author}
  {\bibfnamefont {B}~\bibnamefont {Svistunov}}, \ and\ \bibinfo {author}
  {\bibfnamefont {M}~\bibnamefont {Troyer}},\ }\bibfield  {title} {\enquote
  {\bibinfo {title} {The beliaev technique for a weakly interacting bose
  gas},}\ }\href@noop {} {\bibfield  {journal} {\bibinfo  {journal} {New
  Journal of Physics}\ }\textbf {\bibinfo {volume} {12}},\ \bibinfo {pages}
  {043010} (\bibinfo {year} {2010}{\natexlab{b}})}\BibitemShut {NoStop}%
\bibitem [{\citenamefont {Eckstein}(2009)}]{Eckstein:2009qf}%
  \BibitemOpen
  \bibfield  {author} {\bibinfo {author} {\bibfnamefont {Marin}\ \bibnamefont
  {Eckstein}},\ }\emph {\bibinfo {title} {Nonequilibrium dynamical mean-field
  theory}},\ \href@noop {} {Ph.D. thesis},\ \bibinfo  {school} {Universit{\"a}t
  Augsburg} (\bibinfo {year} {2009})\BibitemShut {NoStop}%
\end{thebibliography}%

\end{document}